\documentclass{llncs}

\usepackage[llncs]{sylvie}
\usepackage{graphicx}
\usepackage{gastex}

\title{Primitivity, Uniform Minimality, and State Complexity of Boolean Operations}
\author{Sylvie Davies}
\institute{University of Waterloo\\
Department of Pure Mathematics\\
\email{sldavies@uwaterloo.ca}}

\newcommand{\tr}[1]{\overset{#1}{\longrightarrow}}

\newcommand{\stc}{\operatorname{sc}}
\newcommand{\any}{\ast}
\newcommand{\qedb}{\hfill$\blacksquare$}
\newcommand{\bm}[1]{{\bf\boldmath #1}}
\renewcommand{\implies}{\Rightarrow}
\renewcommand{\iff}{\Leftrightarrow}

\begin{document}
\maketitle

\begin{abstract} 
A minimal deterministic finite automaton (DFA) is \emph{uniformly minimal} if it always remains minimal when the final state set is replaced by a non-empty proper subset of the state set.
We prove that a permutation DFA is uniformly minimal if and only if its transition monoid is a primitive group.
We use this to study boolean operations on group languages, which are recognized by direct products of permutation DFAs.
A direct product cannot be uniformly minimal, except in the trivial case where one of the DFAs in the product is a one-state DFA. However, non-trivial direct products can satisfy a weaker condition we call \emph{uniform boolean minimality}, where only final state sets used to recognize boolean operations are considered.
We give sufficient conditions for a direct product of two DFAs to be uniformly boolean minimal, which in turn gives sufficient conditions for pairs of group languages to have maximal state complexity under all binary boolean operations (``maximal boolean complexity'').
In the case of permutation DFAs with one final state, we give necessary and sufficient conditions for pairs of group languages to have maximal boolean complexity.
Our results demonstrate a connection between primitive groups and automata with strong minimality properties.
\end{abstract}

\section{Introduction}
Formal definitions are postponed until later.

The state complexity of a regular language is the minimal number of states needed to recognize the language with a deterministic finite automaton.
It is well-known that if $L_m$ and $L'_n$ are regular languages over a common alphabet $\Sig$ with state complexity $m$ and $n$ respectively, then the state complexity of $L_m\cup L'_n$ is at most $mn$, and this bound is tight for all $m,n \ge 2$. The upper bound follows from the standard ``direct product'' automaton construction for recognizing unions of regular languages. Examples which meet the bound were given by Maslov in 1970~\cite{Mas70}, and independently by Yu, Zhuang and Salomaa in 1994~\cite{YZS94}, who noted that the same bound holds for intersection. 

More generally, if $\circ$ is a binary boolean operation on languages over $\Sig$, then $L_m \circ L'_n$ has state complexity at most $mn$, and this bound is tight for all $m,n \ge 2$ if and only if $\circ$ is \e{proper}, that is, not a constant function ($L \circ L' = \emp$ or $L \circ L' = \Sig^*$) or a function that depends on only one argument (for example, $L \circ L' = \Sig^* \setminus L$).
This was proved by Brzozowski in 2009~\cite{Brz09}, who gave examples showing that $mn$ is a tight bound for symmetric difference, and noted that the examples for union and symmetric difference (together with their complements) suffice to prove $mn$ is a tight upper bound for all proper binary boolean operations.

To prove a lower bound on the worst-case state complexity of a regular operation, it suffices to give just one family of examples that meet the bound. Such families are called \e{witnesses}. Witnesses are known for most commonly used unary and binary operations on regular languages. However, there are several directions of research in state complexity which necessitate finding new witnesses for operations that have previously been studied. 
For example, sometimes the first witnesses found for an operation are not optimal in terms of alphabet size, so researchers will look for new witnesses over a smaller alphabet.
When studying the $n$-ary versions of binary operations, such as the union of $n$ languages, or more generally \e{combined operations}~\cite{EGLY09,SSY07}, such as the star of a union of languages, again new witnesses are needed. 
It is also interesting to consider families of languages that are simultaneous witnesses for multiple operations; it is not generally the case that a witness for one operation will work for others. Brzozowski found a family of languages which is a simultaneous witness for reversal, star, concatenation and all binary boolean operations~\cite{Brz13}. 
Each of the problems just mentioned, as well as the fundamental problem of determining the worst-case state complexity of an operation, may also be studied in \e{subclasses} of the regular languages, such as the star-free languages~\cite{BrLi12} or ideal languages~\cite{BJL13}.
Often the known witnesses do not lie in the subclass, so new witnesses must be found. 

In some cases, new witnesses can be found by making slight modifications to known witnesses, but this is not always successful.
Furthermore, this technique does little to advance our understanding of \e{why} particular witnesses work.
For these reasons, it is desirable to have results which describe the \e{general landscape} of witnesses for a particular operation. 
By this we mean results that give \e{necessary conditions} for witnesses, revealing common structural properties that all witnesses share, or \e{sufficient conditions} allowing one to easily generate examples of witnesses or check whether a candidate family is a witness.
For example, Salomaa, Wood and Yu proved that a regular language of state complexity $n$ is a witness for the reversal operation if the transition monoid of its minimal DFA has the maximal possible size $n^n$~\cite{SWY04}; this gives a general sufficient condition for a language to be a witness for reversal.
Ideally, collecting results of this sort would eventually lead to a complete classification of witnesses for commonly used operations. In reality, we suspect the problem of fully classifying witnesses is only tractable in very special cases, but even results which take small steps in this direction can be quite useful and enlightening.

The main inspiration for this work is a paper of Bell, Brzozowski, Moreira, and Reis~\cite{BBMR14}, which considers the following question: for which pairs of languages $(L_m,L'_n)$ (with state complexities $m$ and $n$ respectively) does $L_m \circ L'_n$ reach the maximal state complexity $mn$ for every proper binary boolean operation $\circ$? 
Bell et al.\ give sufficient conditions for this to occur. The conditions are based on the transition monoids of the minimal deterministic automata of $L_m$ and $L'_n$; essentially, if the transition monoids contain the symmetric groups $S_m$ and $S_n$, then ``usually'' (i.e., excluding a known class of counterexamples) the language $L_m \circ L'_n$ will have state complexity $mn$.
We obtain a refinement of this result: we prove that if the transition monoids contain 2-transitive groups, then ``usually'' $L_m \circ L'_n$ has state complexity $mn$ (though our notion of ``usually'' is more restrictive than that of Bell et al.).

We also obtain necessary and sufficient conditions for $L_m \circ L'_n$ to have state complexity $mn$ in the special case where the minimal automata for $L_m$ and $L'_n$ have exactly one final state, and their transition monoids contain a transitive permutation group.
We can view this result as solving a particular special case of the problem of characterizing witnesses for boolean operations.

To obtain these results, we exploit a connection between a certain class of permutation groups called \e{primitive groups}, and the notion of \e{uniformly minimal} automata introduced by Restivo and Vaglica~\cite{ReVa12}.
A minimal deterministic finite automaton (DFA) is \emph{uniformly minimal} if it always remains minimal when the final state set is replaced by a non-empty proper subset of the state set.
For a permutation DFA (that is, a DFA whose transition monoid is a permutation group), uniform minimality is equivalent to primitivity of the transition monoid. 
Although uniform minimality played an important role in the paper of Bell et al., this connection with primitive groups was not used in their paper.
Primitive groups are an important and well-studied class of permutation groups; there are deep results on their structure, and large libraries of primitive groups are available in computer algebra systems such as GAP~\cite{GAP} and Magma~\cite{Magma}. Uniformly minimal DFAs have received comparatively little study; thus this connection has significant implications for the theory of uniformly minimal DFAs. 


The paper is structured as follows. 
Section \ref{sec:def} contains background material needed to understand the paper. 
Section \ref{sec:prum} discusses the relationship between primitive groups and uniformly minimal permutation DFAs.
Section \ref{sec:main} contains our main results on witnesses for the maximal state complexity of boolean operations.
Section \ref{sec:conc} concludes the paper by giving a summary of our results and stating some open problems.

%
%

\section{Definitions and Notation}
\label{sec:def}

For a function $f \co X \ra Y$, we typically write the symbol $f$ to the \e{right} of its arguments. For example, if the image of $x$ under $f$ is $y$, we write $xf = y$. Functions are composed from left to right, and composition is denoted by juxtaposition: if $g \co Y \ra Z$, then $fg$ denotes the composition of $f$ and $g$, and $x(fg) = (xf)g = yg$ is an element of $Z$.

Let $\cP(X)$ denote the \e{power set} of $X$, that is, the set of all subsets of $X$. 
Given $f \co X \ra Y$ we may \e{extend $f$ by union} to obtain a function $f \co \cP(X) \ra \widetilde{Y}$ (where $\widetilde{Y}$ is the closure of $Y$ under union) defined by $Sf = \bigcup_{x \in S} xf$ for $S \subseteq X$. We denote the extension by the same symbol as the original function. Note that for convenience, we often make no distinction between an element of a set and the singleton containing the element; so $x \cup x' = \{x\} \cup \{x'\} = \{x,x'\}$ and $xf = \{x\}f$. 

\subsection{Monoids, Groups and Actions}
A \e{monoid} is a set $M$ equipped with an associative binary operation $\cdot$ and an identity element $e$ such that $m\cdot e = e \cdot m = m$ for all $m \in M$.
Typically we omit the symbol for the operation; so the previous equation could be written as $me = em = m$. 
For $n \ge 1$ we write $m^n$ for the $n$-fold product of $m$ with itself, and define $m^0 = e$ for all $m \in M$.
If for each $m \in M$, there exists $m' \in M$ such that $mm' = m'm = e$, then $M$ is called a \e{group}, and $m'$ is called the \e{inverse} of $M$ and denoted $m^{-1}$.
The \e{order} of an element $g$ of a group is the least integer $n \ge 1$ such that $g^n = e$.

A \e{submonoid} of $M$ is a subset $M' \subseteq M$ which is closed under $\cdot$ and contains the identity $e$ of $M$.
If additionally $M'$ is a group, it is called a \e{subgroup} of $M$; we write $M' \le M$ to mean that $M'$ is a subgroup of $M$. 
Note that we do not allow submonoids or subgroups of $M$ to have an identity element different from that of $M$.
If $x_1,\dotsc,x_k$ are elements of a group $G$, then $\gen{x_1,\dotsc,x_k}$ denotes the \e{group generated by $x_1,\dotsc,x_k$}, the smallest subgroup of $G$ containing $x_1,\dotsc,x_k$.

Let $M$ and $M'$ be monoids with identity elements $e$ and $e'$ respectively.
A \e{homomorphism} from $M$ to $M'$ is a function $\phi \co M \ra M'$ such that $(m_1 m_2)\phi = (m_1)\phi(m_2)\phi$ for all $m_1,m_2 \in M$ and $e\phi = e'$.
A bijective homomorphism is called an \e{isomorphism}, and two monoids are said to be \e{isomorphic} if there exists an isomorphism from one to the other.
We write $M \iso M'$ to mean that $M$ and $M'$ are isomorphic.
If $G$ and $G'$ are groups and $\phi \co G \ra G'$ is a homomorphism, the \e{kernel} of $\phi$ is the set $\ker\phi = \{ g \in G : g\phi = e' \}$, that is, the set of elements of $G$ that map to the identity of $G'$.
If $G$ is a group, $N \le G$, and $gng^{-1} \in N$ for all $g \in G$ and $n \in N$, we say $N$ is a \e{normal subgroup} of $G$. 
A group $G$ is \e{simple} if it has no non-trivial proper normal subgroups, that is, the only normal subgroups of $G$ are $G$ itself and the trivial group (containing just the identity element of $G$).
The kernel of a homomorphism from $G$ to another group is always a normal subgroup of $G$.
We occasionally use the following elementary facts about normal subgroups and homomorphisms:
\bi
\item
If $\phi \co G \ra G'$ is a homomorphism and $\ker\phi$ is the trivial one-element subgroup of $G$, then $\phi$ is injective.
\item
If $\phi \co G \ra G'$ is a \e{surjective} homomorphism and $N$ is a normal subgroup of $G$, then $N\phi$ is a normal subgroup of $G'$.
\ei


A \e{monoid action} of $M$ on a set $X$ is a function $\psi \co X \times M \ra X$ such that $((x,m)\psi,m')\psi = (x,mm')\psi$ and $(x,e)\psi = x$ for all $m,m' \in M$ and $x \in X$. Equivalently, it is a family of functions $m_\psi \co X \ra X$ such that $m_\psi m'_\psi = (mm')_\psi$ for all $m,m' \in M$ and $e_\psi$ is the identity map on $X$.
The map $m_\psi$ is called the \e{action of $m$}.
To simplify the notation, we often omit the action symbol $\psi$ and just write $xm$ instead of $xm_\psi$ or $(x,m)\psi$.
Furthermore, we typically avoid assigning a symbol to the action at all; rather than ``let $\psi$ be a monoid action of $M$ on $X$'' we write ``let $M$ be a monoid acting on $X$'', meaning that $M$ has a specific but nameless action on $X$ associated with it.
If $S \subseteq M$ generates the monoid $M$, a monoid action $\psi$ is completely determined by its values on elements of $S$.
If $M$ is a group, we use the term \e{group action} rather than monoid action.

Let $G$ be a group acting on $X$. 
For $x \in X$, the \e{stabilizer subgroup} or simply \e{stabilizer} of $x$ is the subgroup $\{ g \in G : xg = x\}$ of $G$.
For $S \subseteq X$, the \e{setwise stabilizer} of $S$ is the subgroup $\{ g \in G : Sg = S\}$.
Elements of the setwise stabilizer need not fix every element of $S$; for example, if $1g = 2$ and $2g = 1$ then $g$ is in the setwise stabilizer of $\{1,2\}$.

Let $X$ be a finite set.
A function $t \co X \ra X$ is called a \e{transformation} of $X$.
The set of all transformations of $X$ is a monoid under composition called the \e{full transformation monoid} $T_X$. A submonoid of $T_X$ is called a \e{transformation monoid} on $X$.
The \e{degree} of a transformation monoid on $X$ is the size of $X$.
If $M$ is a transformation monoid on $X$, the monoid action $\psi \co X \times M \ra X$ given by $(x,t)\psi = xt$ for $x \in X$, $t \in M$ is called the \e{natural action} of $M$.
If $X = \{1,\dotsc,n\}$ we write $T_n$ for $T_X$.

A bijective transformation of $X$ is called a \e{permutation} of $X$.
We can describe any particular permutation of $X$ using \e{cycle notation} as follows.
For $x_1,\dotsc,x_k \in X$, we write $(x_1,\dotsc,x_k)$ for the permutation that sends $x_i$ to $x_{i+1}$ for $1 \le i < k$, sends $x_k$ to $x_1$, and fixes all other elements of $X$.
This permutation is called a \e{cycle of length $k$}, or simply a \e{$k$-cycle}.
All permutations that are not cycles can be expressed as a product of cycles.
The identity permutation is denoted by an empty cycle, i.e., ``$()$''.
Cycle notation conflicts with the notation we use for ordered $k$-tuples, but this should not cause confusion. We mainly use cycle notation when giving concrete examples of permutations.

The set of all permutations of $X$ is a subgroup of $T_X$ called the \e{symmetric group} $S_X$.
A subgroup of $S_X$ is called a \e{permutation group} on $X$; this a special type of transformation monoid and we have the same notions of degree and natural action.
The \e{alternating group} $A_X$ is the subgroup of $S_X$ consisting of all permutations that can be expressed as a product of an \e{even number of $2$-cycles}.
If $X = \{1,\dotsc,n\}$ we write $S_n$ for $S_X$ and $A_n$ for $A_X$.

Let $G$ be a group acting on $X$.
We say that the action of $G$ is \e{transitive} or that $G$ \e{acts transitively} on $X$ if for all $x,x' \in X$, there exists $g \in G$ such that $xg = x'$.
We say the action of $G$ is \e{$k$-transitive} or $G$ \e{acts $k$-transitively} on $X$ if for all pairs of $k$-tuples $(x_1,\dotsc,x_k),(x'_1,\dotsc,x'_k) \in X^k$, there exists $g \in G$ such that for $1 \le i \le k$ we have $x_ig = x'_i$; informally, $k$-transitive means ``transitive on $k$-tuples''.

A non-empty set $B \subseteq X$ is called a \e{block} for $G$ if for all $g \in G$, either $Bg \cap B = B$ (equivalently, $Bg = B$) or $Bg \cap B = \emp$.
A block $B$ is \e{trivial} if it is a singleton or the entire set $X$.
We say the action of $G$ is \e{primitive} or that $G$ \e{acts primitively} on $X$ if it is transitive and all of its blocks are trivial.
Equivalently, a transitive group action of $G$ is primitive if for every set $S \subsetneq X$ with at least two elements, there exists $g \in G$ such that $\emp \subsetneq Sg \cap S \subsetneq S$.

If $G$ is a permutation group and the natural action of $G$ is transitive ($k$-transitive, primitive), then we say $G$ is a \e{transitive group} (\e{$k$-transitive group}, \e{primitive group}).
For example, the cyclic group $\gen{(1,2,3,4)} \le S_4$ is a transitive group, since its natural action on $\{1,2,3,4\}$ is transitive.
This terminology can cause confusion, since transitivity, $k$-transitivity and primitivity are properties of \e{actions} and not groups; statements like ``$G$ is transitive'' or ``$G$ is primitive'' are statements about a particular action of $G$ (the natural action) rather than the abstract group itself.
In particular, these properties are not preserved under isomorphism; for example, the group $\gen{(5,6,7,8)} \le S_8$ is not transitive, but it is isomorphic to the transitive group $\gen{(1,2,3,4)} \le S_4$. 

As the notions of transitivity and primitivity are central to this paper, we give numerous examples to illustrate them below.

\bx
\label{ex:cyc-comp}
Consider the group $G = \gen{(1,2,3,4,5,6)} \le S_6$. This group is clearly transitive, since its natural action on $\{1,2,3,4,5,6\}$ is transitive. However, it is imprimitive, since $\{1,3,5\}$ and $\{2,4,6\}$ are non-trivial blocks. Indeed, if we let $a = (1,2,3,4,5,6)$, then $\{1,3,5\}a = \{2,4,6\}$ and $\{2,4,6\}a = \{3,5,1\}$. Hence for all $k \ge 0$, we either have $\{1,3,5\}a^k \cap \{1,3,5\} = \emp$ or $\{1,3,5\}a^k \cap \{1,3,5\} = \{1,3,5\}$, and similarly for $\{2,4,6\}$. One may also verify that $\{1,4\}$, $\{2,5\}$ and $\{3,6\}$ are non-trivial blocks, and that there are no blocks of size 4 or 5. \qedb
\ex

\bx
\label{ex:cyc-prime}
Consider the group $G = \gen{(1,2,3,4,5)} \le S_5$. This group is clearly transitive, and it is also primitive.
To see this, suppose for a contradiction that $B$ is a non-trivial block.
Let $a = (1,2,3,4,5)$ and let $k = |b-b'|$, where $b$ and $b'$ are distinct elements of $B$.
Then $Ba^k \cap B \ne \emp$, so we must have $Ba^k = B$ since $B$ is a block.
Thus for each $i \in B$, we have $ia^k \in Ba^k$, and thus $ia^k \in B$.
Then since $ia^k \in B$, we have $ia^{2k} \in Ba^k$, and thus $ia^{2k} \in B$.
By induction it follows that $\{ ia^{nk} : n \ge 0 \} \subseteq B$.
We claim $\{ia^{nk} : n \ge 0\} = \{1,2,3,4,5\}$, which contradicts the fact that $B$ is a \e{non-trivial} block.
Indeed, for $j \in \{1,2,3,4,5\}$, we have $ia^{nk} = j$ if and only if $i+nk \equiv j \pmod{5}$. Since $5$ is prime and $0 < k < 5$, we see that $k$ is coprime with $5$.
Hence by elementary number theory, there exists $n$ such that $nk \equiv j-i \pmod{5}$ and so $i+nk \equiv i+j-i \equiv j \pmod{5}$ as required.
Hence $j \in \{ia^{nk} : n \ge 0\}$ for all $j \in \{1,2,3,4,5\}$, which proves the claim.
It follows $G$ has no non-trivial blocks, and thus $G$ is primitive. \qedb
\ex

The above argument can be generalized to prove that a cyclic group $G = \gen{(1,2,\dotsc,n)}$ is primitive if and only if $n$ is prime. If $a = (1,2,\dotsc,n)$, then for each divisor $d$ of $n$ and each integer $1 \le i \le n$, we see that $\{ia^{md} : m \ge 0\}$ is a block. In particular, when $n$ is composite, there exists a divisor $d$ with $1 < d < n$, giving rise to a non-trivial block.

\bx
Consider the group $G = \gen{(1,2,3),(4,5,6)} \le S_6$. This group is intransitive, since (for example) it does not contain a permutation mapping $1$ to $4$. Thus it is imprimitive by definition. Alternatively, observe that $\{1,2,3\}$ and $\{4,5,6\}$ are non-trivial blocks for $G$.

Generally an intransitive group will always have non-trivial blocks, but there is one exception: the trivial subgroup of $S_2$ (containing only the identity element). The natural action of this group is clearly not transitive on $\{1,2\}$, but its only blocks are the trivial blocks $\{1\}$, $\{2\}$ and $\{1,2\}$. To avoid dealing with this exception, we require primitive groups to be transitive by definition. \qedb
\ex

The next example shows that we have the following hierarchy of permutation group properties:
\[ \text{(2-transitive) $\implies$ (primitive) $\implies$ (transitive).} \]
These implications do not reverse. Cyclic groups of composite order give examples of transitive imprimitive groups, while cyclic groups of prime order $p \ge 5$ give examples of primitive, non-2-transitive groups. (For example, the group $\gen{(1,2,3,4,5)} \le S_5$ is not 2-transitive on $\{1,2,3,4,5\}$ since nothing maps the pair $(1,2)$ to the pair $(1,3)$.)

\bx
\label{ex:alt}
The alternating group $A_n$ is 2-transitive for $n \ge 4$. Indeed, given $i,i',j,j' \in \{1,\dotsc,n\}$, the permutation $(i,i')(j,j')$ is the product of an even number of 2-cycles, and it maps the pair $(i,j)$ to $(i',j')$.
We claim $A_n$ is also primitive for $n \ge 2$.
To see this, first note that $A_n$ is a cyclic group of prime order for $2 \le n \le 3$.
For $n \ge 4$, suppose for a contradiction that $B$ is a non-trivial block.
Then $B$ has at least two elements $i$ and $j$, but $B$ is not all of $\{1,\dotsc,n\}$.
Choose $k \in \{1,\dotsc,n\} \setminus B$.
Since $A_n$ is 2-transitive, there exists an element $g \in A_n$ which maps the pair $(i,j)$ to $(j,k)$.
Then $Bg \cap B \ne \emp$ (since $Bg$ and $B$ contain $j$), and thus $Bg \cap B = Bg = B$ since $B$ is a block.
But $Bg$ contains $k$ and $B$ does not, which is a contradiction.
Thus all blocks of $A_n$ are trivial, and thus $A_n$ is primitive.
In fact, this argument shows that all 2-transitive groups are primitive. \qedb
\ex

The following fact is immediate from the definitions of transitivity and primitivity, and is frequently useful: if $H$ is a subgroup of $G$ and $H$ is transitive (primitive), then $G$ is also transitive (primitive). For example, the symmetric group $S_n$ is primitive for $n \ge 2$, since it contains the primitive group $A_n$.

So far, we have only looked at cyclic groups and the symmetric and alternating groups.
For our last pair of examples, we consider two subgroups of $S_6$ that are a little more interesting.
\bx
Define $a = (2,4,6)$, $b = (1,5)(2,4)$ and $c = (1,4,5,2)(3,6)$, and let $G = \gen{a,b,c}$.
We claim this group is transitive on $\{1,\dotsc,6\}$.
For $g \in G$ and $i,j \in \{1,\dotsc,6\}$, we will write $i \tr{g} j$ to mean $ig = j$.
Observe that
\[ 1 \tr{c^3} 2 \tr{a^2} 6 \tr{c} 3,\quad 1 \tr{c} 4 \tr{c} 5. \]
Thus for each $i \ne 1$, there is some group element that maps $1$ to $i$.
If $g \in G$ maps $1$ to $i$, then $g^{-1}$ maps $i$ to $1$.
It follows for each $i,j$, there is some element $x$ that maps $i$ to $1$, and another element $y$ that maps $1$ to $j$, giving
\[ i \tr{x} 1 \tr{y} j. \]
Thus $G$ is transitive.
It is also imprimitive, with non-trivial blocks $\{1,3,5\}$ and $\{2,4,6\}$. 
Indeed, we see that 
\[ \{1,3,5\} \tr{a} \{1,3,5\},\;\{1,3,5\} \tr{b} \{5,3,1\},\;\{1,3,5\} \tr{c} \{4,6,2\}. \] 
Hence these sets are non-trivial blocks. 
\qedb
\ex

\newpage

\bx
Define $a = (1,2,3,4,6)$ and $b = (1,2)(3,4)(5,6)$ and let $G = \gen{a,b}$.
It is easy to see that this group is transitive on $\{1,\dotsc,6\}$: just verify that $1$ can be mapped to every other element and use the argument from the previous example.
This group is also primitive. 
To see this, first note that the subgroup $\gen{a}$ acts primitively on $\{1,2,3,4,6\}$, since it is a cyclic group of prime order. Hence a non-trivial block of $G$ cannot be a subset of $\{1,2,3,4,6\}$, so in particular a non-trivial block of $G$ must contain $5$.
Suppose $B$ is a non-trivial block that contains $5$; then $Ba \cap B$ contains $5$ and hence $Ba \cap B = Ba = B$.
Since $B$ is non-trivial, it contains some element $i \ne 5$, and since $Ba = B$ we have $\{i,ia,ia^2,\dotsc,ia^4\} = \{1,2,3,4,6\} \subseteq B$. 
This implies $B = \{1,2,3,4,5,6\}$, and so $B$ is trivial, which is a contradiction.
Thus all blocks of $G$ are trivial, and thus $G$ is primitive. \qedb
\ex

A \e{congruence} of a monoid action of $M$ on $X$ is an equivalence relation on $X$ that is \e{$M$-invariant} in the following sense: if $E$ is an equivalence class, then for all $m \in M$, there exists an equivalence class $E'$ such that $Em \subseteq E'$. 
In other words, if $x$ and $x'$ are equivalent, then $xm$ and $x'm$ are equivalent for all $m \in M$.
The \e{equality congruence} $\{(x,x) : x \in X\}$ in which elements are equivalent only if they are equal, and the \e{full congruence} $X \times X$ in which all elements are equivalent, are called \e{trivial congruences}.
If $M$ is a transformation monoid on $X$, a congruence of the natural action is called an \e{$M$-congruence}.

The notion of congruences leads to an important alternate characterization of primitivity.
In the case of a permutation group $G$ on $X$, notice that for all $S \subseteq X$ and $g \in G$, the set $Sg$ has the same size as $S$. Hence a $G$-congruence has the following property: if $E$ is an equivalence class, then for all $g \in G$, the set $Eg$ is also an equivalence class. In particular, we either have $E \cap Eg = E$ or $E \cap Eg = \emp$ for all $g \in G$; thus the classes of $G$-congruences are blocks. 

In fact, if $G$ is transitive, then every $G$-congruence arises from the blocks of $G$ as follows.
If $B$ is a block for $G$, the \e{block system} corresponding to $B$ is the set $\{ Bg : g \in G\}$.
As the name implies, each set in a block system is also a block for $G$.
Indeed, for all $g' \in G$, we either have $Bgg' \cap Bg = \emp$ or $Bgg' \cap Bg \ne \emp$, and in the latter case, $Bgg'g^{-1} \cap B \ne \emp$.
But $B$ is a block, so this implies $Bgg'g^{-1} = B$ and thus $Bgg' = Bg$.
Thus every set in a block system in a block, so in particular, all distinct sets in a block system are pairwise disjoint.
Furthermore, since $G$ is transitive, each element of $X$ appears in at least one block of the system.
It follows that block systems are partitions of $X$, and thus equivalence relations on $X$.
It is easy to see that block systems are $G$-invariant, and thus are $G$-congruences.

Thus every block gives rise to a block system that is a $G$-congruence, and every $G$-congruence consists of blocks; it follows block systems and $G$-congruences are one and the same if $G$ is a transitive group.
If all $G$-congruences are trivial, then all block systems of $G$ consist only of trivial blocks, and vice versa.
Thus we obtain our alternate characterization of primitivity: a transitive permutation group $G$ on $X$ is primitive if and only if all $G$-congruences are trivial.

Let us revisit some of our earlier examples of primitive and imprimitive groups in the context of this new characterization.
\bx
Consider the imprimitive cyclic group $G = \gen{a = (1,2,3,4,5,6)} \le S_6$ of Example \ref{ex:cyc-comp}.
Put an equivalence relation $\sim$ on $X = \{1,\dotsc,6\}$ by letting $i \sim j$ if $i$ and $j$ have the same parity (odd or even).
Notice that for $i \in X$, the elements $i$ and $ia$ have opposite parity.
Thus $\sim$ is a $G$-congruence, since if $i \sim j$ then $ia \sim ja$, and so if $[i]$ is the equivalence class of $i$ then $[i]a = [ia]$ is also an equivalence class.
In fact, the classes of $\sim$ are just the blocks $\{1,3,5\}$ and $\{2,4,6\}$ we found in Example \ref{ex:cyc-comp}; thus the $G$-congruence $\sim$ corresponds to the block system $\{\{1,3,5\},\{2,4,6\}\}$.
If we define an equivalence relation by $i \sim j$ if $i$ and $j$ are equivalent modulo $3$, we obtain a non-trival $G$-congruence corresponding to the block system $\{\{1,4\},\{2,5\},\{3,6\}\}$.
As for the trivial $G$-congruences, the equality congruence corresponds to the block system $\{\{1\},\{2\},\dotsc,\{6\}\}$ containing the singletons, and the full congruence corresponds to the block system $\{\{1,\dotsc,6\}\}$ that just contains the full set $X$. \qedb
\ex

\bx
Consider the primitive cyclic group $G = \gen{(1,2,3,4,5)} \le S_5$ of Example \ref{ex:cyc-prime}.
With the notion of $G$-congruences, it is much easier to prove that this group is primitive.
Indeed, fix a $G$-congruence on $X$.
By $G$-invariance, all classes of the $G$-congruence must have the same size, say $m$.
If the congruence has $n$ classes, then we have $mn = |X| = 5$.
So $m$ is either $1$ or $5$ since $5$ is prime, which means the classes are either singletons (giving the equality congruence) or the full set $X$ (giving the full congruence).
Thus all $G$-congruences are trivial, and thus $G$ is primitive.
Alternatively, we could make the same argument in terms of block systems, using the fact that all blocks in a system have the same size to show all blocks must be trivial.
This argument actually shows that not only are cyclic groups of prime order primitive, but \e{all transitive groups of prime degree} are primitive (since $|X|$ is the degree of a permutation group on $X$). \qedb
\ex

\subsection{Languages, Automata and State Complexity}
Let $\Sig$ be a finite set. The set of all finite-length sequences of elements of $\Sig$ is called the \e{free monoid} generated by $\Sig$, and is denoted $\Sig^*$. In this context, elements of $\Sig$ are called \e{letters}, and elements of $\Sig^*$ are called \e{words} over $\Sig$. The operation of the free monoid is concatenation of words, and the identity element is the \e{empty word} $\eps$ of length zero.
A set $L \subseteq \Sig^*$ is called a \e{language} over $\Sig$, and $\Sig$ is called the \e{alphabet} of $L$. 

We use the convention that a language $L \subseteq \Sig^*$ is implicitly a \e{pair} $(L,\Sig)$, so for example, the language $\{a,ab\}$ over alphabet $\{a,b\}$ and the language $\{a,ab\}$ over alphabet $\{a,b,c\}$ are distinct.
In particular, two words over different alphabets are necessarily distinct.
This is similar to the convention which views two functions with different codomains as necessarily distinct.

A \e{deterministic finite automaton} (DFA) is a tuple $\cA = (Q,\Sig,\delta,1,F)$ where $Q$ and $\Sig$ are finite sets, $\delta \co Q \times \Sig^* \ra Q$ is a monoid action, $1 \in Q$, and $F \subseteq Q$.
The elements of $Q$ are called \e{states}; the state $1$ is called the \e{initial state} and the states in $F$ are called \e{final states}. The set $\Sig$ is the \e{alphabet} of the automaton. The monoid action $\delta$ is called the \e{transition function}.

Since $\Sig$ generates $\Sig^*$, we may completely specify the action $\delta$ by defining the function $a_\delta \co Q \ra Q$ for each $a \in \Sig$. If $w = a_1\dotsb a_k$ for $a_1,\dotsc,a_k \in \Sig$, then $w_\delta \co Q \ra Q$ is the composition $(a_1)_\delta \dotsb (a_k)_\delta$. The function $\eps_\delta \co Q \ra Q$ is necessarily the identity map.
The monoid $M(\cA) = \{w_\delta \co w \in \Sig^*\}$ is called the \e{transition monoid} of $\cA$; it is a submonoid of $T_Q$ and thus has a natural action on $Q$.
We call the function $w_\delta$ the \e{action of $w$}.
Under our notational conventions, we may write $\delta(p,w) = q$ as $pw_\delta = q$ or simply $pw = q$. We may also extend $w_\delta$ by union and apply it to subsets of the state set: for $X \subseteq Q$ we have $Xw = \{ qw : q \in X\}$.
We also sometimes write $p \tr{w} q$ to mean $pw = q$.

A state $q \in Q$ is \e{reachable from $p \in Q$} if $pw = q$ for some $w$.
Two states $p,q \in Q$ are \e{distinguishable} by $X \subseteq Q$ if there exists $w \in \Sig^*$ such that $pw \in X \iff qw \not\in X$.
We frequently use two special cases of these definitions: A state $q \in Q$ is \e{reachable} if it is reachable from the initial state $1$, and states $p,q \in Q$ are \e{distinguishable} if they are distinguishable by $F$.
We say $\cA$ is \e{accessible} if every state is reachable (from the initial state $1$), and \e{strongly connected} if every state is reachable from every other state.
A state $q \in Q$ is \e{empty} if $qw \not\in F$ for all $w \in \Sig^*$.
In a strongly connected DFA, there exists an empty state if and only if all states are empty.

Consider the following relation on $Q$: two states $p,q \in Q$ are related if and only if they are \e{indistinguishable} by $X \subseteq Q$, that is, for all $w \in \Sig^*$ we have $pw \in X \iff qw \in X$. This is an equivalence relation on $Q$, and in fact it is an $M(\cA)$-congruence. Indeed, if $p$ and $q$ are equivalent, we have $pw \in X \iff qw \in X$ for all $w \in \Sig^*$. So in particular, if we take $w = xy$ for some fixed $x \in \Sig^*$, then $(px)y \in X \iff (qx)y \in X$ for all $y \in \Sig^*$, and thus $px$ and $qx$ are equivalent for all $x \in \Sig^*$.
This congruence is called the \e{indistiguishability congruence of $X$}.

The \e{language recognized by $\cA$} or simply \e{language of $\cA$} is the language $L(\cA) = \{ w \in \Sig^* : 1w \in F\}$ over $\Sig$.
A language which can be recognized by a DFA is called a \e{regular} language.
Two DFAs are \e{equivalent} if they have the same language.
Two DFAs $\cA = (Q,\Sig,\delta,1,F)$ and $\cA' = (Q',\Sig',\delta',1',F')$ with $\Sig = \Sig'$ are \e{isomorphic} if there is a bijection $f \co Q \ra Q'$ such that $1f = 1'$, $Ff = F'$, and $(qa_\delta)f = (qf)a_{\delta'}$ for all $a \in \Sig$; in other words, they are identical up to the naming of the states. In particular, isomorphic DFAs are equivalent.

We say $\cA$ is \e{minimal} if the number of states is minimal among all DFAs equivalent to $\cA$.
It is well-known that for each regular language $L$, all minimal DFAs recognizing $L$ are isomorphic and hence have the same number of states.
The number of states in a minimal DFA for $L$ is called the \e{state complexity} of $L$, and is denoted $\stc(L)$.
A DFA $\cA$ is minimal if and only if all states are reachable and all pairs of states are distinguishable.

Given a binary regular operation $\circ$, the \e{state complexity} of the operation $\circ$ is the following function:
\[ (m,n) \mapsto \max\{ \stc(L_m \circ K_n) : \stc(L_m) = m, \stc(K_n) = n \}. \]
That is, it is the maximal state complexity of the language resulting from the operation, expressed as a function of the state complexities of the operation's arguments. 

For $f \co \bN \times \bN \ra \bN$ and $g \co \bN \times \bN \ra \bN$, we say $f \le g$ if $f(m,n) \le g(m,n)$ for all $(m,n) \in \bN \times \bN$; we say that $f$ is an \e{upper bound} for the state complexity of the operation $\circ$ if $\stc(\circ) \le f$, and a \e{tight upper bound} if $\stc(\circ) = f$.

In the definition of state complexity of operations, we assume that $\circ$ takes two languages over \e{the same alphabet} as arguments. This is justified by our view that words over different alphabets are necessarily distinct; hence, for example, the union of two languages over different alphabets would be a set containing a mixture of words over different alphabets, which is not a language. To perform such an operation, one must first convert the operands to languages over a common alphabet. 
This convention is very common in the literature; however, Brzozowski has recently argued this convention is unnecessary, and in fact leads to incorrect state complexity bounds for operations on languages over different alphabets, since converting the input languages to a common alphabet can change their state complexities~\cite{Brz16}. Brzozowski introduces a distinction between \e{restricted state complexity of operations}, the traditional model in which operands must have the same alphabet, and \e{unrestricted state complexity of operations}, a new model which produces accurate state complexity bounds for operations on languages over different alphabets. 

We use \e{restricted} state complexity in this paper for the following reasons.
First, computing unrestricted state complexity requires using DFAs which have an empty state, and in particular are not strongly connected. In this paper, we mainly study DFAs whose transition monoids are permutation groups, which are always strongly connected. This group-theoretic focus is essential to most of our results. Working with DFAs that are not strongly connected would take us into the realm of semigroup theory, and we are unsure how much of our work would carry over. 
Second, restricted state complexity has been the dominant model of state complexity of operations for many years, while unrestricted state complexity is a recent generalization. Although restricted state complexity gives incorrect results when applied to languages over different alphabets, it is otherwise a correct model.
We have chosen to study the simpler case of restricted state complexity in this paper and leave the more general unrestricted case for potential future work.

\section{Primitive Groups and Uniform Minimality}
\label{sec:prum}
A DFA $\cA = (Q,\Sig,\delta,1,F)$ is called a \e{permutation DFA} if $M(\cA)$ is a permutation group on $Q$. In this case we call $M(\cA)$ the \e{transition group} rather than the transition monoid. The languages recognized by permutation DFAs are called \e{group languages}. 
\bp
\label{prop:trans}
For a permutation DFA $\cA$, the following are equivalent:
\be
\item
\label{p:acc}
$\cA$ is accessible.
\item
\label{p:scon}
$\cA$ is strongly connected.
\item
\label{p:trans}
$M(\cA)$ is transitive.
\ee
\ep
\bpf
\bm{$(\ref{p:acc}) \implies (\ref{p:trans})$}: Since $\cA$ is accessible, for each $q \in Q$ there exists $w_q \in \Sig^*$ such that $1w_q = q$. Since $M(\cA)$ is a group, the element $w_q$ has an inverse, and thus for all $p,q \in Q$ we have $p(w_p)^{-1}w_q = 1w_q = q$. It follows $M(\cA)$ is transitive.

\bm{$(\ref{p:trans}) \implies (\ref{p:scon})$}: Since $M(\cA)$ is transitive, for all $p,q \in Q$ there exists $w \in \Sig^*$ such that $pw = q$. This is precisely saying that $\cA$ is strongly connected.

The last implication \bm{$(\ref{p:scon}) \implies (\ref{p:acc})$} is immediate. \qed
\epf
Note that \bm{$(\ref{p:scon}) \iff (\ref{p:trans})$} holds for arbitrary DFAs, not only permutation DFAs.

Let $\cA = (Q,\Sig,\delta,1,F)$ be a DFA and let $L = L(\cA)$ be its language.
For $S \subseteq Q$, we write $\cA(S)$ for the DFA $\cA = (Q,\Sig,\delta,1,S)$ obtained by replacing the final state set of $\cA$ with $S$.
We say a regular language $L'$ is a \e{cognate} of $L$ if $L' = L(\cA(S))$ for some $S \subseteq Q$.
We say a DFA $\cA'$ is a \e{cognate} of $\cA$ if $\cA' = \cA(S)$ for some $S$; so a language is a cognate of $L$ if and only if it is recognized by a cognate of $\cA$.
If $S = Q$ or $S = \emp$, then $\cA(S)$ is called a \e{trivial} cognate of $\cA$, since $L(\cA(S))$ is either $\Sig^*$ or the empty language $\emp$.

We say $\cA$ is \e{uniformly minimal} if all non-trivial cognates of $\cA$ are minimal. That is, we can reassign the final state set of the DFA in any non-trivial way and the new DFA will always be minimal.
Equivalently, all cognates of $L = L(\cA)$ have the same state complexity $|Q|$.
This definition is essentially restricted to accessible DFAs, since if $\cA$ is not accessible, then not all states are reachable and hence no cognate of $\cA$ can be minimal.

Restivo and Vaglica introduced and studied uniformly minimal DFAs in~\cite{ReVa12g}.
Their notion of uniform minimality is almost the same as ours, except it is restricted to \e{strongly connected} DFAs.
Presumably, Restivo and Vaglica were interested in DFAs that are minimal for every reassignment of \e{initial and final} states; if a DFA is not strongly connected, we can reassign the initial state to obtain a new DFA which is not accessible and hence not minimal.
However, for strongly connected DFAs, the choice of initial state has no effect on minimality since every state is reachable from each possible choice of initial state.
Hence we lose nothing by fixing an initial state and generalizing to accessible DFAs.

\brs
Restivo and Vaglica also studied uniformly minimal DFAs in~\cite{ReVa12}, but they used different terminology. They used the term ``almost uniformly minimal'' for the notion discussed above, and used ``uniformly minimal'' for a stronger condition that can only be met by \e{incomplete} DFAs (which we do not discuss in this paper).
\ers

A DFA $\cA$ is called \e{simple} if all $M(\cA)$-congruences are trivial.
\'{E}sik proved the following result for strongly connected DFAs~\cite[Proposition 1]{ReVa12g}. The same proof works for accessible DFAs. 
\bp
\label{prop:prum}
An accessible DFA $\cA$ is uniformly minimal if and only if it is simple.
\ep
\bpf
Suppose $\cA$ is simple, that is, all $M(\cA)$-congruences are trivial. Then in particular, for every $S \subseteq Q$, the indistiguishability congruence of $S$ is trivial. If the indistiguishability congruence for $S$ is the equality relation, then each state lies in its own class, so all pairs of states are distinguishable. 
Since $\cA$ is accessible, all states are reachable, and hence $\cA$ is minimal. If the indistinguishability congruence for $S$ is the full relation, then all states are indistinguishable. But final states are always distinguishable from non-final states, so this can only happen if all states are final ($S = Q$) or all states are non-final ($S = \emp$). Hence if $\emp \subsetneq S \subsetneq Q$, then $\cA(S)$ is minimal, so it follows that $\cA$ is uniformly minimal.

Conversely, suppose $\cA$ is not simple, and there exists a non-trivial $M(\cA)$-congruence. Then this congruence has a class $E$ which has at least two elements, but is not all of $Q$. Let $E$ be the final state set of $\cA$ and let $p,q \in E$. For all $w \in \Sig^*$, the states $pw$ and $qw$ both lie in the set $Ew$, which is contained in some congruence class $E'$. If $E' = E$, then we have $pw, qw \in E$. If $E' \cap E = \emp$, then we have $pw,qw \not\in E$. Thus for all $w \in \Sig^*$, we have $pw \in E \iff qw \in E$, and so $p$ and $q$ are not distinguishable by $E$. Hence $\cA$ is not uniformly minimal, since $\cA(E)$ is not minimal. \qed
\epf

In the special case of permutation DFAs, we have:
\bc
\label{cor:prum}
An accessible permutation DFA $\cA$ is uniformly minimal if and only if $M(\cA)$ is primitive.
\ec
\bpf
If $\cA$ is uniformly minimal, then it is simple, so all $M(\cA)$-congruences are trivial. Now, recall that a group $G$ is primitive if and only if all $G$-congruences are trivial. Since $M(\cA)$ is a group, we see that $M(\cA)$ is primitive.

Conversely, if $M(\cA)$ is primitive, then all $M(\cA)$-congruences are trivial. Hence $\cA$ is simple and hence uniformly minimal. \qed
\epf
Note that both implications in Corollary \ref{cor:prum} are vacuously true if $\cA$ is not accessible: $\cA$ cannot be uniformly minimal, and $M(\cA)$ cannot be transitive and thus cannot be primitive. 
Thus one can technically omit the accessible assumption.

It seems this relationship between primitivity and minimality has been overlooked until recently.
Primitive groups have seen increasing application in automata theory over the past decade, particularly in connection with the classical \e{synchronization problem} for DFAs; for a survey of such work see~\cite{ACS15}.
The connection between simple DFAs and primitive groups was recently noted by Almeida and Rodaro~\cite{AlRo16}. 
However, primitive groups are not mentioned in Restivo and Vaglica's work on uniformly minimal DFAs, nor in any other work on DFA minimality that we are aware of.

The wealth of results on primitive groups makes Corollary \ref{cor:prum} quite useful for studying and constructing uniformly minimal DFAs. For example, we can use this corollary to easily prove that for each $n \ge 2$, there exists a uniformly minimal DFA with $n$ states.
Restivo and Vaglica proved this using a rather complicated construction~\cite[Theorem 3]{ReVa12}.
\bp
For each $n \ge 2$, there exists a uniformly minimal permutation DFA with $n$ states.
\ep
\newpage
\bpf
The symmetric group $S_n$ is primitive for all $n \ge 2$, and clearly for each $n \ge 2$ there exists an $n$-state DFA with transition group $S_n$. For example, let $\{g_1,\dotsc,g_k\}$ be a generating set of the symmetric group and let $\cA$ be a DFA with states $\{1,\dotsc,n\}$, alphabet $\Sig = \{a_1,\dotsc,a_k\}$, and transition function $\delta$ with $(a_i)_\delta = g_i$ for $1 \le i \le k$. In fact we can use a binary alphabet, 
since $S_n$ has generating sets of size two for all $n \ge 2$.
\qed
\epf

This proof illustrates a technique that is very useful for producing examples of DFAs. If we have a generating set for a transformation monoid, we can construct a DFA which has that monoid as its transition monoid.
\bx
\label{ex:altdfa}
Let $\cA$ be the DFA with alphabet $\{a,b\}$ defined as follows. 
\bi
\item
The states are $\{1,2,3,4\}$, the initial state is $1$, and the final states are $\{3,4\}$.
\item
The transformations are the permutations $a = (2,3,4)$ and $b = (1,2)(3,4)$.
\ei
More formally, we mean that the transition function $\delta$ of $\cA$ is given by $a_\delta = (2,3,4)$ and $b_\delta = (1,2)(3,4)$. However, we will generally be brief when describing DFAs, as above.

The permutations $(1,2)(3,4)$ and $(2,3,4)$ generate the alternating group $A_4$. Thus the transition group of $\cA$ is $A_4$. We saw in Example \ref{ex:alt} that $A_4$ is transitive and primitive. Hence by Proposition \ref{prop:trans}, $\cA$ is strongly connected, and by Corollary \ref{cor:prum}, $\cA$ is uniformly minimal.

A state diagram of $\cA$ is given in Figure \ref{fig:altdfa}. We can see from the diagram that $\cA$ is indeed strongly connected. It is tedious, but possible to verify that $\cA$ is uniformly minimal by checking that it is minimal with respect to every non-empty, proper subset of $\{1,2,3,4\}$. \qedb
\ex

\begin{figure}[h]
\unitlength 11pt
\begin{center}\begin{picture}(18,4)(0,0)
\gasset{Nh=2,Nw=2,Nmr=1,ELdist=0.3,loopdiam=1}

\node(1)(0,0){$1$}
\node(2)(6,0){$2$}
\node(3)(12,0){$3$}
\node(4)(18,0){$4$}
\imark(1)
\rmark(3)\rmark(4)

\drawloop(1){$a$}
\drawedge[curvedepth=1](1,2){$b$}
\drawedge[curvedepth=1](2,1){$b$}
\drawedge[curvedepth=1](4,3){$b$}
\drawedge(2,3){$a$}
\drawedge[curvedepth=1](3,4){$a,b$}
\drawedge[curvedepth=-4](4,2){$a$}
\end{picture}\end{center}
\caption{Uniformly minimal DFA $\cA$ of Example \ref{ex:altdfa}.}
\label{fig:altdfa}
\end{figure}
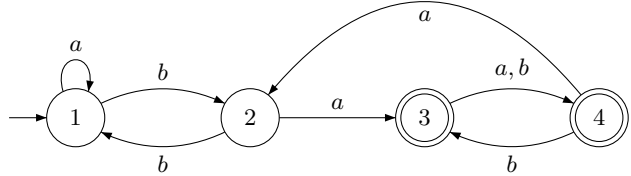

\begin{figure}[h]
\unitlength 11pt
\begin{center}\begin{picture}(20,4)(0,0)
\gasset{Nh=2,Nw=2,Nmr=1,ELdist=0.3,loopdiam=1}

\node(1)(0,0){$1$}
\node(2)(4,0){$2$}
\node(3)(8,0){$3$}
\node(4)(12,0){$4$}
\node(5)(16,0){$5$}
\node(6)(20,0){$6$}
\imark(1)
\rmark(1)
\rmark(3)
\rmark(5)

\drawedge(1,2){$a$}
\drawedge(2,3){$a$}
\drawedge(3,4){$a$}
\drawedge(4,5){$a$}
\drawedge(5,6){$a$}
\drawedge[curvedepth=-3](6,1){$a$}
\end{picture}\end{center}
\caption{Non-minimal DFA $\cA$ of Example \ref{ex:cycdfa} with an imprimitive transition group.}
\label{fig:cycdfa}
\end{figure}
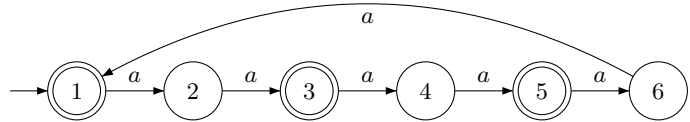

\bx
\label{ex:cycdfa}
Let $\cA$ be the DFA with alphabet $\{a\}$, states $\{1,\dotsc,6\}$, initial state $1$, final states $F = \{1,3,5\}$ and $a = (1,2,3,4,5,6)$. A diagram is in Figure \ref{fig:cycdfa}.

The transition group $G$ of $\cA$ is the cyclic group of order six, which is imprimitive.
We saw in Example \ref{ex:cyc-comp} that $F = \{1,3,5\}$ is a block for this group.
Hence for all $k$, we either have $Fa^k = F$ or $Fa^k \cap F = \emp$.
Thus if $i,j \in F$, then for all $k$ we have $ia^k \in F \iff ja^k \in F$.
This means all pairs of states in $F$ are indistinguishable by $F$, and hence $\cA$ is not minimal.

This argument actually shows that whenever $F$ is a non-trivial block of $G$, the DFA $\cA$ is not minimal. In fact, this also holds whenever $F$ is a union of non-trivial blocks of $G$ (see Lemma \ref{lem:prum} below). 

Note that if we construct a DFA from a cyclic group of \e{prime} order, we get a uniformly minimal DFA, since cyclic groups of prime order are primitive.
\qedb
\ex

There exist many infinite families of primitive groups, and hence of uniformly minimal permutation DFAs. However, there are infinitely many positive integers $n$ for which the only primitive groups of degree $n$ are $S_n$ and $A_n$~\cite[pg.\ 66]{DiMo96}. Hence other infinite families of primitive groups cannot be used to construct $n$-state uniformly minimal DFAs for every $n$, unless we ``fill in the gaps'' with symmetric or alternating groups.

\brs
Steinberg has extended the notion of primitivity to transformation monoids~\cite{Ste10}. 
Steinberg defines a transformation monoid $M$ to be \e{primitive} if there are no non-trivial $M$-congruences. Under this definition, an accessible DFA $\cA$ is uniformly minimal if and only if the transition monoid $M(\cA)$ is primitive. However, we have not investigated whether any of our other results that hold for primitive groups are also true for primitive monoids.
\ers

We close this section with a technical lemma that generalizes Proposition \ref{prop:prum}. If $M$ is a transformation monoid on $X$ and $S \subseteq X$, we say that $S$ is \e{saturated} by an $M$-congruence if it is a union of classes of the $M$-congruence.
\bl
\label{lem:prum}
An accessible DFA $\cA$ with $\emp \subsetneq F \subsetneq Q$ is minimal if and only if there is no non-trivial $M(\cA)$-congruence that saturates $F$.
\el
It follows that if all $M(\cA)$-congruences are trivial, then $\cA$ is uniformly minimal. Conversely, if there is a non-trivial $M(\cA)$-congruence, then it saturates its own congruence classes and at least one class is a proper non-empty subset of $Q$, and thus $\cA$ is not uniformly minimal. Hence this indeed generalizes Proposition \ref{prop:prum}.
\bpf
We prove the contrapositive: $\cA$ is not minimal if and only if there exists a non-trivial $M(\cA)$-congruence that saturates $F$.

Suppose $\cA$ is not minimal. Then the indistinguishability congruence of $F$ is a non-trivial $M(\cA)$-congruence, since at least two states are indistinguishable.
Suppose there is an indistinguishability class $E$ that is neither contained in $F$ nor disjoint from $F$.
Then there exist $p,q \in E$ such that $p \in F$ and $q \not\in F$.
But then $p$ and $q$ are distinguishable by $F$, which cannot happen since $E$ is an indistinguishability class.
Thus for each indistinguishability class $[q]$, we have $[q] \subseteq F$ or $[q] \cap F = \emp$.
Then we have $F = \bigcup_{f \in F} [f]$, so $F$ is saturated by its indistinguishability congruence.

Conversely, let $E_1,\dotsc,E_k \subseteq Q$ be the congruence classes of a non-trivial $M(\cA)$-congruence that saturates $F$.
Choose a congruence class $E_i$ of size at least two. 
Then for all $w \in \Sig^*$ we have $E_iw \subseteq E_j$ for some $j$.
Since $F$ is a union of congruence classes, either $E_j \subseteq F$ or $E_j \cap F = \emp$.
Hence for $p,q \in E_i$ and all $w \in \Sig^*$, we have $pw \in F \iff E_j \subseteq F \iff E_iw \subseteq F \iff qw \in F$.
It follows that states in $E_i$ are indistinguishable, and thus $\cA$ is not minimal. \qed
\epf
In the special case of permutation DFAs, this has a useful consequence.
\bc
\label{cor:1fstate}
Let $\cA$ be a permutation DFA.
If $|F|=1$ or $|F| = |Q|-1$, then $\cA$ is minimal if and only if it is accessible.
\ec
\bpf
Recall that if $G$ is a transitive permutation group and $E$ and $E'$ are classes of a $G$-congruence, then $|E| = |E'|$. 
It follows that if $|F| = 1$, then a non-trivial $M(\cA)$-congruence cannot saturate $F$ since all the congruence classes have size at least two. Furthermore, an $M(\cA)$-congruence saturates $F$ if and only if it saturates $Q \setminus F$, and if $|F| = |Q|-1$ then a non-trivial $M(\cA)$-congruence cannot saturate the set $Q \setminus F$ of size one. Hence if $\cA$ is accessible, it is minimal by Lemma \ref{lem:prum}.
On the other hand, if $\cA$ is not accessible, it cannot be minimal. \qed
\epf

\section{Main Results}
\label{sec:main}

Throughout this section, $\cA = (Q,\Sig,\delta,1,F)$ and $\cA' = (Q',\Sig',\delta',1',F')$ are minimal DFAs with a common alphabet $\Sig = \Sig'$.
The languages of $\cA$ and $\cA'$ are $L$ and $L'$, and the transition monoids are $M$ and $M'$, respectively.
For $w \in \Sig^*$, we write $w$ for $w_\delta \in M$ and $w'$ for $w_{\delta'} \in M'$.
Sometimes we will assume $\cA$ and $\cA'$ are permutation DFAs, and then we will use $G$ and $G'$ for the transition groups rather than $M$ and $M'$.

\subsection{Direct Products and Boolean Operations}
The \e{direct product} of $\cA$ and $\cA'$ is the DFA $\cA \times \cA'$ with state set $Q \times Q'$, alphabet $\Sig$, transitions $(q,q') \tr{a} (qa,q'a')$ for each $(q,q') \in Q \times Q'$ and $a \in \Sig$, initial state $(i,i')$, and an unspecified set of final states.
By assigning particular sets of final states to $\cA \times \cA'$ as described below, we can recognize the languages resulting from arbitrary binary boolean operations on $L$ and $L'$.


Fix a function $\circ \co \{0,1\}^2 \times \{0,1\}$; these are called \e{binary boolean functions}.
For a set $S$, let $\chi_S \co S \ra \{0,1\}$ denote the \e{characteristic function} of $S$, defined by $\chi_S(x) = 1$ if $x \in S$ and $\chi_S(x) = 0$ otherwise.
We can think of $\chi_S(x)$ as giving the ``truth value'' of the proposition ``\,$x \in S$\,'', where 0 is false and 1 is true.
Now for $F \subseteq Q$ and $F' \subseteq Q'$, define $F \circ F' = \{(q,q') \in Q \times Q' : \chi_F(q) \circ \chi_{F'}(q') = 1\}$.
Then $\cA \times \cA'$ with final states $F \circ F'$ recognizes the language $L \circ L'$ defined by
\[ L \circ L' = \{ w \in \Sig^* : \chi_L(w) \circ \chi_{L'}(w) = 1 \}. \]
For example, if $\circ \co \{0,1\}^2 \ra \{0,1\}$ is the ``logical or'' function, then $L \circ L' = L \cup L'$, since $w \in L \circ L'$ if $w \in L$ or $w \in L'$. Similarly, the ``logical and'' function gives the intersection $L \cap L'$.

We say that a boolean function (and the associated boolean operation on languages) is \e{proper} if its output depends on both of its arguments.
For example, $u \circ v = 1-u$ (giving $L \circ L' = \Sig^* \setminus L$) only depends on the first argument, and $u \circ v = 0$ (giving with $L \circ L' = \emp$) depends on neither argument, so they are not proper. If $L$ and $L'$ have state complexity $m$ and $n$ respectively, then the improper binary boolean operations have state complexity $1$ (if they are constant), state complexity $m$ (if they depend only on the first operand), or state complexity $n$ (if they depend only on the second operand). There are 16 binary boolean operations in total, and one may easily verify that 10 of them are proper.

If $\cA$ and $\cA'$ have $m$ and $n$ states respectively, $\cA \times \cA'$ has $mn$ states.
Hence every proper binary boolean operation has state complexity bounded by $mn$. 
It is well-known that this bound is tight for general regular languages, and it remains tight for regular group languages. In fact, the original witnesses for union given by Maslov~\cite{Mas70} and Yu et al.~\cite{YZS94} are group languages, and we will demonstrate later (in Example \ref{ex:wit}) that these languages are also witnesses for all other proper binary boolean operations.
We will say the pair $(L,L')$ has \e{maximal boolean complexity} if $\stc(L \circ L') = \stc(L)\stc(L')$ for all proper binary boolean operations $\circ$.

Suppose that $\emp \subsetneq F \subsetneq Q$ and $\emp \subsetneq F' \subsetneq Q'$.
We say a subset of $Q \times Q'$ is \e{$(F,F')$-compatible} if it is equal to $F \circ F'$ for some proper binary boolean operation $\circ$.
Notice that $(L,L')$ has maximal boolean complexity if and only if $\cA \times \cA'$ is minimal for every $(F,F')$-compatible subset of $Q \times Q'$.
We disallow $F = \emp$ and $F = Q$ since then $\cA \times \cA'$ is minimal for $F \circ F'$ only if $|Q| = 1$ and $L = \emp$ or $L = \Sig^*$; these cases are uninteresting. Similarly, we exclude $F' = \emp$ and $F' = Q'$.
We say the pair $(\cA,\cA')$ (or the direct product $\cA \times \cA'$) is \e{uniformly boolean minimal} if for every pair of sets $(S,S')$ with $\emp \subsetneq S \subsetneq Q$ and $\emp \subsetneq S' \subsetneq Q'$ and every $(S,S')$-compatible set $S \circ S'$, the DFA $(\cA \times \cA')(S \circ S')$ is minimal. 
In other words, if every pair of \e{cognates} of $L$ and $L'$ has maximal boolean complexity.

We give an example of a pair of DFAs that are not uniformly boolean minimal, as well as a pair of DFAs that are.

\bx
\label{ex:symdiff2}
Define two DFAs over alphabet $\Sig = \{a,b,c\}$ as follows:
\bi
\item $\cA$ has state set $Q = \{1,2\}$, initial state $1$, final state set $F = \{1\}$, and transformations $a = b = (1,2)$, $c = ()$.
\item $\cA'$ has state set $Q' = \{1,2\}$, initial state $1$, final state set $F' = \{1\}$, and transformations $a' = c' = (1,2)$, $b' = ()$.
\ei
We show that $(L(\cA),L(\cA'))$ does not have maximal boolean complexity, and thus $\cA \times \cA'$ is not uniformly boolean minimal. 
\newpage

\begin{figure}[h]
\unitlength 11pt
\begin{center}\begin{picture}(24,12)(0,0)
\gasset{Nh=2,Nw=2,Nmr=1,ELdist=0.3,loopdiam=1}

\node(1)(0,9){$1$}
\node(2)(6,9){$2$}
\node(3)(14,12){$1,1$}
\node(4)(26,12){$1,2$}
\node(5)(0,3){$1$}
\node(6)(6,3){$2$}
\node(7)(14,0){$2,1$}
\node(8)(26,0){$2,2$}
\imark(1)\imark(5)\imark(3)
\rmark(1)\rmark(5)\rmark(4)\rmark(7)

\drawloop(1){$c$}
\drawloop(2){$c$}
\drawedge[curvedepth=1](1,2){$a,b$}
\drawedge[curvedepth=1](2,1){$a,b$}

\drawloop(5){$b$}
\drawloop(6){$b$}
\drawedge[curvedepth=1](5,6){$a,c$}
\drawedge[curvedepth=1](6,5){$a,c$}

\drawedge[curvedepth=1](3,4){$c$}
\drawedge[curvedepth=1](4,3){$c$}
\drawedge[curvedepth=1](7,8){$c$}
\drawedge[curvedepth=1](8,7){$c$}

\drawedge[curvedepth=1](3,7){$b$}
\drawedge[curvedepth=1](7,3){$b$}
\drawedge[curvedepth=1](4,8){$b$}
\drawedge[curvedepth=1](8,4){$b$}

\drawedge[curvedepth=1](3,8){$a$}
\drawedge[curvedepth=1](8,3){$a$}
\drawedge[curvedepth=1](4,7){$a$}
\drawedge[curvedepth=1](7,4){$a$}

\end{picture}\end{center}
\caption{DFAs $\cA$, $\cA'$ and $\cA \times \cA'$ of Example \ref{ex:symdiff2}. The final state set of $\cA \times \cA'$ is chosen so that $\cA \times \cA'$ recognizes the symmetric difference of the languages of $\cA$ and $\cA'$.}
\label{fig:symdiff2}
\end{figure}
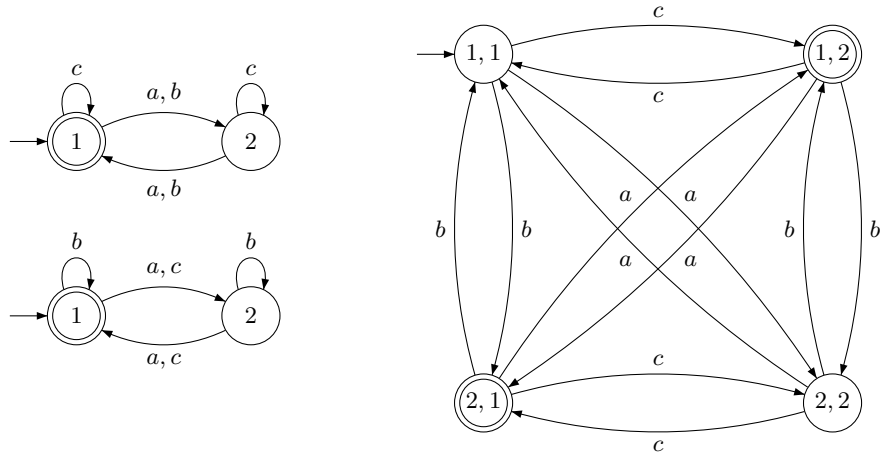

To see this, consider the symmetric difference operator $\oplus$, arising from the ``exclusive or'' boolean function: for $u,v \in \{0,1\}$, the ``exclusive or'' $u \oplus v$ is zero if $u = v$ and one if $u \ne v$. The corresponding operation on languages over $\Sig$ is
\[ L \oplus L' = \{ w \in \Sig^*: w \in L \text{ or } w \in L' \text{, but not both}\} = (L \setminus L') \cup (L' \setminus L). \]
The final state set that makes $\cA \times \cA'$ recognize $L(\cA) \oplus L(\cA')$ is:
\[ F \oplus F' = \{ (q,q') \in Q \times Q : q \in F \text{ or } q' \in F' \text{, but not both}\} = \{(1,2),(2,1)\}. \]
State diagrams of the DFAs $\cA$, $\cA'$ and $(\cA \times \cA')(F \oplus F')$ are shown in Figure \ref{fig:symdiff2}. Notice that $(\cA \times \cA')(F \oplus F')$ is not minimal: the states $(1,2)$ and $(2,1)$ cannot be distinguished. Since $F \oplus F'$ is an $(F,F')$-compatible set, it follows that $(L(\cA),L(\cA'))$ does not have maximal boolean complexity, and that $\cA \times \cA'$ is not uniformly boolean minimal. \qedb
\ex

\bx
\label{ex:ubm}
Define two DFAs over alphabet $\Sig = \{a,b\}$ as follows:
\bi
\item $\cA$ has state set $Q = \{1,2\}$ and transformations $a = (1,2)$, $b = ()$.
\item $\cA'$ has state set $Q' = \{1,2,3\}$ and transformations $a' = (1,2)$, $b' = (1,2,3)$.
\ei
The initial and final states are not important for this example.

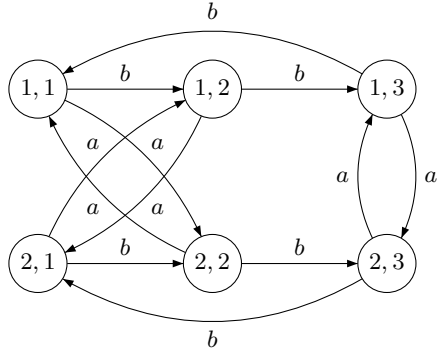
\begin{figure}[h]
\unitlength 11pt
\begin{center}\begin{picture}(12,10)(0,-1)
\gasset{Nh=2,Nw=2,Nmr=1,ELdist=0.3,loopdiam=1}

\node(1)(0,8){$1,1$}
\node(2)(6,8){$1,2$}
\node(3)(12,8){$1,3$}
\node(4)(0,2){$2,1$}
\node(5)(6,2){$2,2$}
\node(6)(12,2){$2,3$}

\drawedge[curvedepth=1](1,5){$a$}
\drawedge[curvedepth=1](5,1){$a$}
\drawedge[curvedepth=1](2,4){$a$}
\drawedge[curvedepth=1](4,2){$a$}
\drawedge[curvedepth=1](3,6){$a$}
\drawedge[curvedepth=1](6,3){$a$}

\drawedge(1,2){$b$}
\drawedge(2,3){$b$}
\drawedge[curvedepth=-2,ELdist=-1](3,1){$b$}
\drawedge(4,5){$b$}
\drawedge(5,6){$b$}
\drawedge[curvedepth=2](6,4){$b$}
\end{picture}\end{center}
\caption{Uniformly boolean minimal DFA $\cA \times \cA'$ of Example \ref{ex:ubm}.}
\label{fig:ubm}
\end{figure}

The direct product $\cA \times \cA'$ is shown in Figure \ref{fig:ubm}.
Notice that the transition group of $\cA'$ is $S_3$.
We will see much later (Corollary \ref{cor:dissimilar}) that this implies $\cA \times \cA'$ is uniformly boolean minimal.

Note that $\cA \times \cA'$ is not uniformly minimal; for example, it is not minimal with respect to the final state set $\{(1,1),(1,2),(1,3)\}$. If $|Q|,|Q'| \ge 2$, a direct product DFA with state set $Q \times Q'$ can never be uniformly minimal. In particular, it cannot be minimal for final state sets of the form $S \times Q'$ (``unions of rows'') or $Q \times S'$ (``unions of columns''). However, the definition of uniform boolean minimality excludes these sets. \qedb
\ex

Bell, Brzozowski, Moreira and Reis found sufficient conditions for a pair of DFAs to be uniformly boolean minimal~\cite{BBMR14}. However, these conditions require that the transition monoids of the DFAs \e{contain the symmetric group}, in the sense that they contain every permutation of the DFA's state set.
In particular, for permutation DFAs, these conditions only apply when the transition group \e{is} the symmetric group on the state set.
We obtain more general sufficient conditions for uniform boolean minimality in permutation DFAs, which apply to a larger class of transition groups.
Additionally, we show that DFAs whose transition monoids contain 2-transitive groups ``usually'' meet these conditions, up to some technical assumptions we will state later.

We also obtain necessary and sufficient conditions for a pair of languages $(L,L')$ to have maximal boolean complexity in the special case where $L$ and $L'$ are recognized by permutation DFAs $\cA$ and $\cA'$ with exactly one final state. In this special case, it turns out $(L,L')$ has maximal boolean complexity if and only if $\cA \times \cA'$ is accessible. We give several group-theoretic conditions and a graph-theoretic condition that are equivalent to $\cA \times \cA'$ being accessible.

We begin with a proposition which characterizes $(F,F')$-compatible subsets. If $Q$ is the state set of a DFA and $S \subseteq Q$, write $\ol{S}$ for $Q \setminus S$. Similarly, if $L$ is a language over $\Sig$, write $\ol{L}$ for $\Sig^* \setminus L$.
\bp
\label{prop:compat}
Let $\emp \subsetneq F \subsetneq Q$ and $\emp \subsetneq F' \subsetneq Q'$.
A subset of $Q \times Q'$ is $(F,F')$-compatible if and only it is equal to one of the following sets:
\be[(a)]
\item
$F \times F'$ (corresponding to $L \cap L'$).
\item
$F \times \ol{F'}$ (corresponding to $L \cap \ol{L'} = L \setminus L'$).
\item
$\ol{F} \times F'$ (corresponding to $\ol{L} \cap L' = L' \setminus L$).
\item
$\ol{F} \times \ol{F'}$ (corresponding to $\ol{L} \cap \ol{L'} = \ol{L \cup L}$).
\item
$(F \times \ol{F'}) \cup (\ol{F} \times F')$ (corresponding to symmetric difference $(L \setminus L') \cup (L' \setminus L)$).
\item
The complement $(Q \times Q') \setminus S$, where $S$ is one of the above sets.
\ee
\ep
\bpf
Let $\circ$ be a proper binary boolean function.
Let $k$ be the number of pairs $(u,v) \in \{0,1\} \times \{0,1\}$ such that $u \circ v = 1$.

\bm{Case 1 ($k = 1$)}:
If $k = 1$, then there is a unique pair $(u,v)$ such that $u \circ v = 1$.
Hence $F \circ F' = \{ (q,q') : \chi_F(q) = u \text{ and } \chi_{F'}(q') = v\}$.
Consider possible values for $(u,v)$:
\bi
\item
If $(u,v) = (0,0)$ then $F \circ F' = \ol{F} \times \ol{F'}$.
\item
If $(u,v) = (0,1)$ then $F \circ F' = \ol{F} \times F'$.
\item
If $(u,v) = (1,0)$ then $F \circ F' = F \times \ol{F'}$.
\item
If $(u,v) = (1,1)$ then $F \circ F' = F \times F'$.
\ei
Hence $F \circ F'$ is a set of type (a), (b), (c) or (d).

\bm{Case 2 ($k = 2$)}:
If $k = 2$, there are exactly two pairs $(u,v)$ and $(u',v')$ such that $u \circ v = u' \circ v' = 1$.
We claim that $u \ne u'$ and $v \ne v'$.
To see this, suppose $u = u'$.
Then we must have $v \ne v'$, or else the pairs are not distinct.
Thus $\{v,v'\} = \{0,1\}$ and it follows $u \circ 0 = u \circ 1 = 1$.
Hence $\circ$ only depends on the value of the first argument, which contradicts the fact that $\circ$ is proper.
By a symmetric argument, we cannot have $v = v'$.
Now, observe that $(q,q')$ is in $F \circ F'$ if and only if
\[
\chi_F(q) = u \text{ and } \chi_{F'}(q') = v
 \text{ or } 
\chi_F(q) = u' \text{ and } \chi_{F'}(q') = v'. \]
Suppose $(u,v) = (1,0)$. Then we necessarily have $(u',v') = (0,1)$ and we get
\[ F \circ F' = (F \times \ol{F'}) \cup (\ol{F} \times F').  \]
If $(u,v) = (0,1)$, then $(u',v') = (1,0)$ and we get the same set.
If $(u,v) = (1,1)$ or $(u,v) = (0,0)$, then we get
\[ F \circ F' = (F \times F') \cup (\ol{F} \times \ol{F'}).  \]
But this is simply the complement of the previous set.
So we either get a set of type (e) or the complement of such a set, which is type (f). 

\bm{Case 3 ($k = 3$)}:
If $k = 3$, then there is a unique pair $(u,v)$ such that $u \circ v = 0$.
Hence $F \circ F'$ is the \e{complement} of a set of type (a), (b), (c) or (d), that is, a set of type (f).

This proves that every $(F,F')$-compatible set, that is, every set of the form $F \circ F'$ where $\circ$ is a proper binary boolean function, has one of the given forms (a)--(f). Conversely, if we are given sets $F$ and $F'$ and a set $X \subseteq Q \times Q'$ with one of the forms (a)--(f), the proof shows how to construct a proper binary boolean function $\circ$ such that $X = F \circ F'$. It follows $X$ is $(F,F')$-compatible if and only if it has one of the forms (a)--(f). \qed
\epf

\subsection{Accessibility of $\cA \times \cA'$}
In this section, we consider the problem of determining when $\cA \times \cA'$ is accessible.
This is essential for proving that $\cA \times \cA'$ is minimal for a certain final state set, and also an interesting question in its own right.
Recall that by Proposition \ref{prop:trans}, the DFA $\cA \times \cA'$ is strongly connected if and only if the transition monoid of $\cA \times \cA'$ is transitive.
Furthermore, if $\cA \times \cA'$ is a permutation DFA, then it is accessible if and only if its transition group is transitive.
The following proposition describes the structure of the transition monoid of $\cA \times \cA'$. 
\bp
\label{prop:monoid}
Let $M_{\times}$ denote the transition monoid of $\cA \times \cA'$.
\be
\item
\label{mo:iso}
$M_{\times}$ is isomorphic to the submonoid of $M \times M'$
generated by $\{(a,a') : a \in \Sig\}$.
We often identify $M_\times$ with this submonoid.
\item
\label{mo:sd}
The projections $\pi \co M_{\times} \ra M$ and $\pi' \co M_{\times} \ra M'$ given by $(w,w')\pi = w$ and $(w,w')\pi' = w'$ are surjective. 
\item
\label{mo:gr}
If $M$ and $M'$ are groups, then $M_{\times}$ is a group.
\ee
\ep
\bpf
\bm{(\ref{mo:iso})}:
Write $w_\times$ for $w_{\cA \times \cA'} \in M_\times$.
Consider the map $\phi \co M_\times \ra M \times M'$ given by $w_\times \mapsto (w,w')$.
This map is clearly a monoid homomorphism.
Furthermore, if $(x,x') = (y,y')$ then $qx = qy$ and $q'x' = q'y'$ for all $q \in Q$ and $q' \in Q'$, and thus in $M_\times$ we have $(q,q')x_{\times} = (q,q')y_{\times}$ for all $(q,q') \in Q \times Q'$.
Hence $x_\times = y_\times$ whenever $x_\times \phi = y_\times \phi$, and it follows that $\phi$ is injective.

Since $\phi$ is injective, $(M_\times)\phi$ is a finite monoid of the same size as $M_\times$.
It follows $\phi$ is \e{bijective} when viewed as homomorphism between $M_\times$ and $(M_\times)\phi$, and thus $\phi$ is an isomorphism between these monoids.
Since $\{a_{\times} : a \in \Sig\}$ generates $M_\times$, we see that $\{(a,a') : a \in \Sig\}$ generates $(M_\times)\phi$.
Hence we have $M_\times \iso (M_\times)\phi = \gen{(a,a') : a \in \Sig}$ as required.

\bm{(\ref{mo:sd})}:
Fix $w \in M$.
Then for the element $(w,w') \in M_\times$ we have $(w,w')\pi = w$. 
Hence $\pi$ maps surjectively onto $M$.
Similarly, $\pi'$ maps surjectively onto $M'$.

\bm{(\ref{mo:gr})}:
Since $M_\times$ is a monoid, it suffices to show every element of $M_\times$ has an inverse. 
Recall that the identity elements of $M$ and $M'$ are $\eps$ and $\eps'$ respectively.
For $(w,w') \in M_\times$, pick $m$ and $n$ such that $w^m = \eps$ in $M$ and $(w')^n = \eps'$ in $M'$. 
This is possible since $M$ and $M'$ are finite groups.
We have $(w^{mn-1},(w')^{mn-1}) \in M_\times$, and $(w,w')(w^{mn-1},(w')^{mn-1}) = (w^{mn-1},(w')^{mn-1})(w,w') = (w^{mn},(w')^{mn}) = (\eps,\eps')$, the identity of $M_{\times}$.
Thus $(w^{mn-1},(w')^{mn-1})$ is an inverse of $(w,w')$. \qed
\epf
Recall that for permutation DFAs $\cA$ and $\cA'$, we denote the transition group of $\cA$ by $G$ and the transition group of $\cA'$ by $G'$; in this case we will also write $G_{\times}$ for the transition group of $\cA \times \cA'$.
If $\cA$ and $\cA'$ are permutation DFAs, the transitivity of $G_\times$ is a necessary and sufficient condition for all states of $\cA \times \cA'$ to be reachable.
However, the structure of $G_\times$ can be difficult to understand.
Hence we will derive a simpler characterization of transitivity that depends only on properties of $G$ and $G'$.

Suppose that $\cA$ and $\cA'$ are permutation DFAs.
Consider the subgroup $\ker\pi \le G_{\times}$.
It contains all $(w,w') \in G_{\times}$ such that $w$ is the identity in $G$.
View $Q \times Q'$ as a grid, where elements of $Q$ are ``row indices'' and elements of $Q'$ are ``column indices''.
Then $\ker\pi$ consists of the elements of $G_\times$ which fix all row indices.
Hence we define $R = \ker\pi$ and call $R$ the \e{full row stabilizer}.
Similarly, $C = \ker\pi'$ fixes all column indices and we call it the \e{full column stabilizer}.
Both of these subgroups are normal, since they are kernels of homomorphisms.

Fix $q \in Q$ and $q' \in Q'$.
Let $(q,\any)$ denote the set $\{(q,i') \in Q \times Q' : i' \in Q'\}$, that is, the ``$q$-th row'' of $Q \times Q'$. Similarly, let $(\any,q') = \{(i,q') \in Q \times Q' : i \in Q\}$ denote the ``$q'$-th column'' of $Q \times Q'$.
Let $R_q \le G_{\times}$ be the setwise stabilizer of $(q,\any)$ and let $C_{q'} \le G_{\times}$ be the setwise stabilizer of $(\any,q')$. 
We call the subgroups $R_q$ the \e{single row stabilizers} and the subgroups $C_{q'}$ the \e{single column stabilizers}.
The full row stabilizer $R$ is the intersection of all single row stabilizers, and hence is a subgroup of each $R_q$; the analogous fact holds for $C$.

We now give necessary and sufficient conditions for $\cA \times \cA'$ to be transitive in the case where $\cA$ and $\cA'$ are permutation DFAs. 
\bl
\label{lem:trans}
Let $\cA$ and $\cA'$ be permutation DFAs. The following are equivalent:
\be
\item
\label{lt:con}
$\cA \times \cA'$ is accessible.
\item
\label{lt:all}
$G$ and $G'$ are transitive and for all $q \in Q$ and $q' \in Q'$, the subgroups $R_{q}\pi' \le G'$ and $C_{q'}\pi \le G$ are transitive.
\item
\label{lt:rowcol}
$G$ is transitive and $R_{q}\pi' \le G'$ is transitive for some $q \in Q$, or $G'$ is transitive and $C_{q'}\pi \le G$ is transitive for some $q' \in Q'$.
\item
\label{lt:times}
$G_{\times}$ is transitive.
\ee
\el
\bpf
Since $\cA \times \cA'$ is a permutation DFA, we see that 
{\boldmath $(\ref{lt:con}) \iff (\ref{lt:times})$}.
Also, the implication
{\boldmath $(\ref{lt:all}) \implies (\ref{lt:rowcol})$}
is immediate.

\bm{$(\ref{lt:rowcol}) \implies (\ref{lt:times})$}:
Fix $(i,i'),(j,j') \in Q \times Q'$. 
Suppose that $R_q\pi'$ is transitive for some $q \in Q$; the case where some $C_{q'}\pi$ is transitive is symmetric.
\bi
\item
Since $G$ is transitive, there exists $x \in G$ such that $ix = q$.
Let $k' \in Q'$ be the element such that $i'x' = k'$.
Then $(i,i') \tr{x} (q,k')$.
\item
Since $G$ is transitive, there exists $y \in G$ such that $qy = j$ in $G$.
\item
Since $R_q\pi' \le G'$ is transitive on $Q'$, there exists $z' \in R_q\pi'$ such that $k'z' = j'(y')^{-1}$.
Since $(z,z') \in R_q$, we have $qz = q$.
Hence $(q,k') \tr{z} (q,j'(y')^{-1})$.
\ei
It follows that 
\[ (i,i') \tr{x} (q,k') \tr{z} (q,j'(y')^{-1}) \tr{y} (qy,j') = (j,j'). \] 
Thus $G_\times$ is transitive on $Q \times Q'$, since for all elements $(i,i'),(j,j') \in Q \times Q'$, there exists an element of $G_\times$ that maps one to the other.

\bm{$(\ref{lt:times}) \implies (\ref{lt:all})$}:
If $G_\times$ is transitive, then for all $q,i,j \in Q$ and $q',i',j' \in Q'$, there exist $x,y \in \Sig^*$ such that
\[(q,i') \tr{x} (qx,i'x') =(q,j') \text{ and } (i,q') \tr{y} (iy,q'y') = (j,q'). \]
Thus, we see that:
\bi
\item
Since $qx = q$, we have $(x,x') \in R_q$, and since $q'y' = q'$, we have $(y,y') \in C_{q'}$.
\item
For all $i,j \in Q$, there exists a word $y \in C_{q'}\pi \le G$ that maps $i$ to $j$.
\item
For all $i',j' \in Q'$, there exists a word $x' \in R_{q}\pi' \le G'$ that maps $i'$ to $j'$.
\ei
Hence for all $q \in Q$ and $q' \in Q'$, we see that $R_q\pi'$ is transitive on $Q'$ and $C_{q'}\pi$ is transitive on $Q$.
Since $C_{q'}\pi \le G$ and $R_q\pi' \le G'$, it follows that $G$ and $G'$ are transitive.

This establishes a cycle of implications 
{\boldmath $(\ref{lt:all}) \implies (\ref{lt:rowcol}) \implies (\ref{lt:times}) \implies (\ref{lt:all})$}.
Since we also have 
{\boldmath$(\ref{lt:con}) \iff (\ref{lt:times})$},
all the statements are equivalent. \qed
\epf

This lemma reduces the problem of checking transitivity of $G_\times$ to just checking the the transitivity of a row stabilizer on the column indices, or of a column stabilizer on the row indices. 
The following proposition gives a graph-theoretic interpretation of this idea, which may be easier to understand and apply. This graph-theoretic condition for accessibility of $\cA \times \cA'$ can easily be proved without appeal to group theory, but for illustrative purposes we will connect it with condition (\ref{lt:rowcol}) of Lemma \ref{lem:trans}.

\bp
\label{prop:graph}
Let $\cA$ and $\cA'$ be permutation DFAs. The following statements are equivalent:
\be
\item
\label{pt:graph}
$\cA$ is accessible and there exists $q \in Q$ such that all states in $(q,\any)$ are reachable, or $\cA'$ is accessible and there exists $q' \in Q'$ such that all states in $(\any,q')$ are reachable.
\item
\label{pt:rowcol}
$G$ is transitive and $R_{q}\pi' \le G'$ is transitive for some $q \in Q$, or $G'$ is transitive and $C_{q'}\pi \le G$ is transitive for some $q' \in Q'$.
\ee
\ep
\bpf
\bm{$(\ref{pt:graph}) \implies (\ref{pt:rowcol})$}:
Suppose $\cA$ is accessible, and there exists $q \in Q$ such that all states in $(q,\any)$ are reachable. Since $\cA$ is an accessible permutation DFA, $G$ is transitive on $Q$.

To see that $R_{q}\pi' \le G'$ is transitive on $Q'$, fix $i',j' \in Q'$.
Since all states in $(q,\any)$ are reachable from the initial state $(1,1')$ of $\cA \times \cA'$, there is some word $x \in G_\times$ such that $(1,1') \tr{x} (q,i')$.
Also, there is some $y \in G_\times$ such that $(1,1') \tr{y} (q,j')$.
Since $\cA \times \cA'$ is a permutation DFA, there is some $z \in \Sig^*$ such that $z = x^{-1}$.
Thus we have
\[ (q,i') \tr{z} (1,1') \tr{y} (q,j'). \]
We see that $zy$ maps $q$ to itself, so $z'y' \in R_{q}\pi'$. 
Hence $R_{q}\pi'$ is transitive on $Q'$.

By a symmetric argument, if $\cA'$ is accessible and there exists $q' \in Q'$ such that all states in $(\any,q')$ are reachable, it follows that $G'$ is transitive on $Q'$ and $C_{q'}\pi \le G$ is transitive on $Q$.

\bm{$(\ref{pt:rowcol}) \implies (\ref{pt:graph})$}:
Suppose $G$ is transitive and $R_{q}\pi' \le G'$ is transitive for some $q \in Q$.
Since $G$ is transitive, $\cA$ is accessible.
In particular, there exists $w \in \Sig^*$ such that $1w = q$, and it follows that $(1,1') \tr{w} (q,1'w')$ in $\cA \times \cA'$.
Since $R_{q}\pi'$ is transitive on $Q'$, for all $q' \in Q$ there exists $x \in R_{q}\pi'$ such that $(q,1'w') \tr{x} (q,q')$.
Hence every state in $(q,\any)$ is reachable.
In the case where $G'$ and some $C_{q'}\pi$ are transitive, we can use a symmetric argument. \qed
\epf

We now prove one of our main results, which gives necessary and sufficient conditions for pairs of group languages recognized by DFAs with exactly one final state to have maximal boolean complexity. 
\newpage
\bt
\label{thm:1fstate}
Suppose $|Q| \ge 3$ or $|Q'| \ge 3$. Let $\cA$ and $\cA'$ be permutation DFAs with exactly one final state.
Then the following are equivalent:
\be
\item
\label{t1f:acc}
$\cA \times \cA'$ is accessible.
\item
\label{t1f:all}
For all proper binary boolean operations $\circ$, the language $L \circ L'$ has maximal state complexity.
That is, $(L,L')$ has maximal boolean complexity.
\item
\label{t1f:one}
There exists a proper binary boolean operation $\circ$ such that the language $L \circ L'$ has maximal complexity.
\ee
\et
Determining whether $\cA \times \cA'$ is accessible can be difficult in general. Perhaps the easiest method is to use the graph-theoretic condition of Proposition \ref{prop:graph}, which states that assuming $\cA$ and $\cA'$ are accessible, the direct product $\cA \times \cA'$ is accessible if either of the following holds.
\bi
\item
There exists a row $(q,\any)$ such that all states in $(q,\any)$ are reachable.
\item
There exists a column $(\any,q')$ such that all states in $(\any,q')$ are reachable.
\ei
This reduces the problem to just checking reachability for a single row or column.

The group-theoretic conditions of Lemma \ref{lem:trans} may also be used, but they are perhaps harder to understand.
Much later in the paper (Section \ref{sec:dissimilar}) we will use these to obtain simpler group-theoretic conditions for accessibility of $\cA \times \cA'$. 
In particular, provided that $\cA$ and $\cA'$ are both accessible, and also satisfy an additional criterion called \e{dissimilarity} (which is usually easy to check), we have:
\bi
\item
If $G$ or $G'$ is a transitive simple group, then $\cA \times \cA'$ is accessible.
\item
If $G$ or $G'$ is a primitive group, then $\cA \times \cA'$ is accessible.
\ei

\bpf[Theorem \ref{thm:1fstate}]
The only difficult implication here is \bm{$(\ref{t1f:acc}) \implies (\ref{t1f:all})$}.
Suppose $\cA \times \cA'$ is accessible; we want to show that $L \circ L'$ has maximal state complexity for every proper binary boolean operation $\circ$. That is, we want to show that all pairs of states of $\cA \times \cA'$ are distinguishable by each $(F,F')$-compatible subset of $Q \times Q'$.

Note that since $\cA \times \cA'$ is accessible, by Lemma \ref{lem:trans} we know that $G_{\times}$ is transitive and that $R_q\pi' \le G'$ and $C_{q'}\pi \le G$ are transitive for all $q \in Q$ and $q' \in Q'$. What this means is:
\bi
\item
For every pair of states $(p,p')$ and $(q,q')$ of $\cA \times \cA'$, there exists a word $w \in \Sig^*$ such that $(p,p') \tr{w} (q,q')$. 
(Transitivity of $G_{\times}$)
\item
Fix a state $q \in Q$. For every pair of states $i',j' \in Q'$, there exists a word $w \in \Sig^*$ such that $(q,i') \tr{w} (q,j')$. (Transitivity of $R_{q}\pi'$)
\item
Fix a state $q' \in Q'$. For every pair of states $i,j \in Q$, there exists a word $w \in \Sig^*$ such that $(i,q') \tr{w} (j,q')$. (Transitivity of $C_{q'}\pi$)
\ei
We will use these facts repeatedly throughout the proof. 

Let $F = \{f\}$ and $F' = \{f'\}$, so that $F \times F' = \{(f,f')\}$.
Let $(p,p')$ and $(q,q')$ be distinct states of $\cA \times \cA'$ that we wish to distinguish.
We will show these states are distinguishable with respect to each type of set described in Proposition \ref{prop:compat}. 

We only need to consider types (a) through (e), since sets of type (f) are just complements of sets of types (a) through (e), and two states are distinguishable by a set $X$ if and only if they are distinguishable by the complement of $X$.

\bm{Case 1 (States in the same row or same column):}
Suppose $p = q$, that is, both states $(p,p')$ and $(q,q')$ are in the same row.
Then we necessarily have $p' \ne q'$, since the states are distinct.
\bi
\item
By transitivity of $G_{\times}$, for all $r \in Q$ there exists $w \in \Sig^*$ such that $(p,p') \tr{w} (r,f')$.
\item
Since $p = q$ and $p' \ne q'$, we have $(q,q') \tr{w} (r,s)$ for some $s \ne f'$. (Since $w'$ is a permutation, it must map $p'$ and $q'$ to different states.)
\ei
If \bm{$r \in F$}, we have $(r,f') \in F \times F'$ and $(r,s) \in F \times \ol{F'}$.
Hence we can distinguish the states if the final state set is 
\bm{$F \times F'$}, \bm{$F \times \ol{F'}$}, or \bm{$(F \times \ol{F'}) \cup (\ol{F} \times F')$}.\\
If \bm{$r \not\in F$}, we have $(r,f') \in \ol{F} \times F'$ and $(r,s) \in \ol{F} \times \ol{F'}$.
Hence we can distinguish the states if the final state set is \bm{$\ol{F} \times F'$} or \bm{$\ol{F} \times \ol{F'}$}.

This covers all the possible sets of final states.
If $p \ne q$ and $p' = q'$ (that is, the states are in the same column) we can use a symmetric argument.

\bm{Case 2 (States in different rows and different columns):}
Assume $p \ne q$ and $p' \ne q'$. 
We consider each possible set of final states in turn.

\bm{$F \times F'$}: Here $\cA \times \cA'$ has exactly one final state $(f,f')$, so it is minimal by Corollary \ref{cor:1fstate}.

\bm{$F \times \ol{F'}$}: We make a few observations:
\bi
\item
By transitivity of $G_\times$, there exists $w \in \Sig^*$ such that $(p,p') \tr{w} (f,f')$.
\item
Since $p \ne q$, $p' \ne q'$ and $w$ is a permutation, we must have $qw \ne f$ and $q'w' \ne f'$.
\item
Since $C_{f'}\pi$ is transitive, there exists $x \in \Sig^*$ such $(qw,f') \tr{x} (f,f')$.
\ei
\item
It follows that 
\[ (p,p') \tr{w} (f,f') \tr{x} (fx,f'),\quad (q,q') \tr{w} (qw,q'w') \tr{x} (f,q'w'x'). \]
\bi
\item
Since $qwx = f$, $qw \ne f$ and $x$ is a permutation, we have $fx \ne f$. It follows that $(fx,f') \in \ol{F} \times F'$.
\item
Since $f'x' = f'$, $q'w' \ne f'$ and $x'$ is a permutation, we have $q'w'x' \ne f'$. It follows that $(f,q'w'x') \in F \times \ol{F'}$.
\ei
Hence $wx$ maps $(p,p')$ to a non-final state and $(q,q')$ to a final state.
Thus we have distinguished the two states.

\bm{$\ol{F} \times F'$}: We can use a symmetric argument to the previous case.

\bm{$\ol{F} \times \ol{F'}$}: 
As in the case of $F \times \ol{F'}$, pick $w$ such that $(p,p') \tr{w} (f,f')$. 
Then $(q,q') \tr{w} (qw,q'w')$, which is in $\ol{F} \times \ol{F'}$ since $qw \ne f$ and $q'w' \ne f'$. Thus $w$ sends $(q,q')$ to a final state.
But $(p,p') \tr{w} (f,f')$ is non-final, so we have distinguished the states.

\newpage
\bm{$(F \times \ol{F'}) \cup (\ol{F} \times F')$}: 
This is the most complicated case.
\bi
\item
By transitivity of $G_\times$, there exists $u \in \Sig^*$ such that $(p,p') \tr{u} (f,r')$, where $r' \ne f'$.
\item
We have $(f,r') \in F \times \ol{F'}$, so $u$ sends $(p,p')$ to a final state.
If $(q,q') \tr{u} (qu,q'u')$ is non-final, then $u$ distinguishes the states, so we may assume without loss of generality that it is final.
\item
We cannot have $qu = f$, since $p \ne q$ and $pu = f$. 
Thus $qu \in \ol{F}$.
Since $(qu,q'u')$ is final we therefore must have $q'u' \in F'$, that is, $q'u' = f'$.
\ei
Define $r = qu$; now we have reduced the problem to distinguishing two states of the forms $(f,r')$ and $(r,f')$, with $r \ne f$ and $r' \ne f'$.

Suppose $|Q| \ge 3$; if we only have $|Q'| \ge 3$ we can use a symmetric argument to the argument below.
\bi
\item
Since $C_{f'}\pi$ is transitive and $|Q| \ge 3$, there is a word $v \in \Sig^*$ such that 
$(f,f') \tr{v} (s,f')$ for some $s \not\in \{r,f\}$.
\item
It follows that $(f,r') \tr{v} (s,r'v')$, where $r'v' \ne f'$.
\item
The state $(s,r'v')$ is in $\ol{F} \times \ol{F'}$, and thus is non-final.
If $(r,f') \tr{v} (rv,f')$ is final, then $v$ distinguishes $(f,r')$ and $(r,f')$. 
Hence we may assume without loss of generality that $(rv,f')$ is non-final.
\item
A non-final state either lies in $F \times F'$ or $\ol{F} \times \ol{F'}$.
Since $f' \in F'$, we must have $(rv,f') \in F \times F'$. But then $rv = f$.
\ei
Thus we have
\[ (p,p') \tr{u} (f,r') \tr{v} (s,r'v'),\quad(q,q') \tr{u} (r,f') \tr{v} (f,f'). \]
Now, apply $v$ again to both states.
\bi
\item
Since $s \ne r$ and $rv = f$, we have $sv \ne f$.
\item
Since $f'v' = f'$ and $r' \ne f'$, we have $r'v' \ne f'$ and $r'v'v' \ne f'$.
\item
It follows that $(s,r'v') \tr{v} (sv,r'v'v') \in \ol{F} \times \ol{F'}$, and thus is non-final.
\item
However, recall that $fv = s$ and $s \ne f$; thus $(f,f') \tr{v} (s,f')$ is in $\ol{F} \times F'$.
\ei
Hence $(p,p')$ and $(q,q')$ are distinguished by $uv^2$. 

We have shown that all pairs of states of $\cA \times \cA'$ are distinguishable by all $(F,F')$-compatible sets of final states, and so this proves \bm{$(\ref{t1f:acc}) \implies (\ref{t1f:all})$}.

The implication \bm{$(\ref{t1f:all}) \implies (\ref{t1f:one})$} is immediate. For \bm{$(\ref{t1f:one}) \implies (\ref{t1f:acc})$}, just note that for each proper binary boolean operation $\circ$, the language $L \circ L'$ is recognized by $(\cA \times \cA')(X)$ for some set of final states $X$.
If $(\cA \times \cA')(X)$ is not accessible, then it cannot be minimal and thus $L \circ L'$ cannot have maximal state complexity. 
\qed
\epf
Note that Example \ref{ex:symdiff2} gives a pair of languages recognized by two-state permutation DFAs which have maximal complexity for intersection, but not symmetric difference (see also~\cite[Example 2]{BBMR14}). Hence in the previous theorem, it was necessary to assume that at least one DFA has three or more states. 
\newpage

Note that Theorem \ref{thm:1fstate} also holds in the following cases:
\bi
\item
$\cA$ and $\cA'$ both have exactly one \e{non-final} state.
\item
$\cA$ has exactly one final state and $\cA'$ has exactly one non-final state.
\item
$\cA$ has exactly one non-final state and $\cA'$ has exactly one final state.
\ei
The same arguments we gave in Theorem \ref{thm:1fstate} can be used in the above three cases, but the role of each argument is changed.
For example, consider the case where $\cA$ has one final state and $\cA'$ has one non-final state.
Let $F = \{f\}$ and let $\ol{F'} = Q' \setminus F' = \{q'\}$.
We can use the same arguments as in the original proof of Theorem \ref{thm:1fstate}, except wherever $F'$ appears we substitute $\ol{F'}$.
So for example, we deal with the case of $F \times \ol{F'} = \{(f,q')\}$ by appealing to Corollary \ref{cor:1fstate}, just like we did for $F \times F'$ in the original proof.
This works because distinguishability arguments are the same whether we distinguish with respect to a set of final states or a set of non-final states.

We now apply Theorem \ref{thm:1fstate} to show that the original witnesses for the maximal state complexity of union (found by Maslow and later by Yu, Zhuang and Salomaa) are in fact witnesses for all proper binary boolean operations.

\bx
\label{ex:wit}
In~\cite{Mas70}, Maslov defined two families of DFAs over alphabet $\{0,1\}$ as follows, and claimed that the languages they recognize are witnesses for union.
The DFA $\cA$ has states $\{S_0,\dotsc,S_{m-1}\}$ with $S_0$ initial and $S_{m-1}$ final, and the transitions are given by $S_i0 = S_i$, $S_{i}1 = S_{i+1}$ for $i \ne m-1$, and $S_{m-1}1 = S_0$. The DFA $\cB$ has states $\{P_0,\dotsc,P_{n-1}\}$ with $P_0$ initial and $P_{n-1}$ final, and the transitions are given by $P_i1 = P_i$, $P_i0 = P_{i+1}$ for $i \ne n-1$, and $P_{n-1}0 = P_0$.
It is easy to see that $\cA \times \cB$ is accessible: the state $(S_i,P_j)$ can be reached from $(S_0,P_0)$ via the word $1^i0^j$.
Furthermore, $\cA$ and $\cB$ are permutation DFAs: the symbol $0$ acts as the identity permutation in $\cA$ and as a cyclic permutation of the states in $\cB$, while $1$ acts as a cyclic permutation in $\cA$ and the identity in $\cB$.
They also have exactly one final state.
So in fact, for $m,n \ge 3$, the pair of languages $(L(\cA),L(\cB))$ has maximal boolean complexity by Theorem \ref{thm:1fstate}.
That is, Maslov's languages are witnesses for all proper binary boolean operations, in addition to union.

Yu, Zhuang and Salomaa gave a different family of witnesses in~\cite{YZS94}.
For a word $w \in \Sig^*$ and $a \in \Sig$, let $|w|_a$ denote the number of occurrences of the letter $a$ in $w$.
Yu et al.\ defined languages $L_m = \{ w \in \{a,b\}^* : |w|_a \equiv 0 \pmod{m}\}$ and $L_n = \{ w \in \{a,b\}^* : |w|_b \equiv 0 \pmod{n}\}$, then proved that $L_m \cap L_n$ and $\ol{L_m} \cup \ol{L_n}$ both have state complexity $mn$.
In fact, $(L_m,L_n)$ has maximal boolean complexity for $m,n \ge 3$.
Indeed, one may verify that the minimal DFA of $L_m$ has $m$ states, with the initial and final states equal; the letter $a$ acts as a cyclic permutation of the state set, and the letter $b$ acts as the identity. The minimal DFA of $L_n$ is similar, except there are $n$ states, $b$ is the cyclic permutation, and $a$ is the identity.
These DFAs are almost identical to the DFAs defined by Maslov, except with a different choice of final state. This does not change the fact that they are permutation DFAs with one final state and an accessible direct product, and hence meet the conditions of Theorem \ref{thm:1fstate}. \qedb
\ex

%

\subsection{Uniform Boolean Minimality}
We now give sufficient conditions for a pair of permutation DFAs $(\cA,\cA')$ to be uniformly boolean minimal.
Just as with uniform minimality, primitive groups play an important role in our conditions for uniform boolean minimality.
To state our conditions, we need some new notation.

For $p,q \in Q$ and $p',q' \in Q'$, we write $R_{p,q}$ for the setwise stabilizer of $(p,\any) \cup (q,\any)$, and $C_{p',q'}$ for the setwise stabilizer of $(\any,p') \cup (\any,q')$. 
If $p = q$ then $R_{p,q} = R_p = R_q$, and similarly if $p' = q'$ then $C_{p',q'} = C_{p'} = C_{q'}$.
We call these subgroups \e{double row stabilizers} and \e{double column stabilizers}.
Under this definition, single row stabilizers are special cases of double row stabilizers, and similarly for column stabilizers.

Note that $R_p$ and $R_q$ are not necessarily subgroups of $R_{p,q}$, nor the other way around: the group $R_p$ might contain elements that map $q$ to some state $r \not\in \{p,q\}$, while the group $R_{p,q}$ might contain elements that swap $p$ and $q$.
However, the full row stabilizer $R$ is a common subgroup of $R_p$, $R_q$ and $R_{p,q}$.
The analogous facts hold for column stabilizers.

\bl
\label{lem:boolprim}
Suppose $\cA$ and $\cA'$ are permutation DFAs. 
\be
\item
\label{bp:row}
If $|Q| \ge 3$, the group $G$ is primitive, and $R_{p,q}\pi' \le G'$ is primitive for all $p,q \in Q$, then $\cA \times \cA'$ is uniformly boolean minimal.
\item
\label{bp:col}
If $|Q'| \ge 3$, the group $G'$ is primitive, and $C_{p',q'}\pi \le G$ is primitive for all $p',q' \in Q'$, then $\cA \times \cA'$ is uniformly boolean minimal.
\ee
\el
Note that we do not require $p \ne q$ or $p' \ne q'$, so in case (\ref{bp:row}) both the single and double row stabilizers must be primitive, and in case (\ref{bp:col}) both the single and double column stabilizers must be primitive.

\bpf
Suppose that the conditions of (\ref{bp:row}) hold, that is, $|Q| \ge 3$, $G$ is primitive and for all $p,q \in Q$, the subgroup $R_{p,q}\pi' \le G'$ is primitive. The case where the conditions of (\ref{bp:col}) hold is symmetric.
We want to show that for every pair of sets $\emp \subsetneq S \subsetneq Q$ and $\emp \subsetneq S' \subsetneq Q'$ and each $(S,S')$-compatible subset $X \subseteq Q \times Q'$, the DFA $(\cA \times \cA')(X)$ is minimal.

It suffices to consider the cases where $X = S \times S'$ and where $X = (S \times S') \cup (\ol{S} \times \ol{S'})$.
It may seem that this would only cover sets of type (a) and complements of sets of type (e) from Proposition \ref{prop:compat}.
However, these two cases actually cover all possible types of $(S,S')$-compatible sets.

To see this, consider a set $S \times \ol{S'}$ of type (b).
If we prove that $(\cA \times \cA')(T \times T')$ is minimal for \e{all} pairs of sets $\emp \subsetneq T \subsetneq Q$ and $\emp \subsetneq T' \subsetneq Q'$, then in particular we can take $T = S$ and $T' = \ol{S'}$ to show that $(\cA \times \cA')(S \times \ol{S'})$ is covered.
A similar argument works for sets of type (c) and (d) .

Now note that if $(\cA \times \cA')(X)$ is minimal, then $(\cA \times \cA')(\ol{X})$ is also minimal, so we also cover sets of type (e) and (f).
So all types of sets from Proposition \ref{prop:compat} are covered by just looking at the cases $X = S \times S'$ and $X = (S \times S') \cup (\ol{S} \times \ol{S'})$.

Let $(i,i')$ and $(j,j')$ be distinct states of $\cA \times \cA'$; we will show they are distinguishable with respect to $X$.

\bm{Case 1 (States in the same row)}:
Suppose $i = j$, that is, the states are in the same row.
Since the states are distinct, we have $i' \ne j'$.
\bi
\item
Since $G$ is primitive, it is transitive, and thus there exists a word $w \in G$ that maps $i = j$ to some element $s \in S$.
Thus $(i,i') \tr{w} (s,i'w')$ and $(j,j') \tr{w} (s,j'w')$.
\item
Suppose $w$ does \e{not} distinguish $(i,i')$ and $(j,j')$ with respect to $X$.
Then $(s,i'w') \in X \iff (s,j'w') \in X$.
\item
Since $R_{s}\pi'$ is primitive, all permutation DFAs with state set $Q'$ and transition group $R_{s}\pi'$ are uniformly minimal by Corollary \ref{cor:prum}.
It follows there exists $x' \in R_{s}\pi'$ that distinguishes $i'w'$ and $j'w'$ with respect to $S'$.
\ei
We have:
\[ (i,i') \tr{w} (s,i'w') \tr{x} (s,i'w'x'),\quad (j,j') \tr{w} (s,j'w') \tr{x} (s,j'w'x'). \]
\bi
\item
Since $x'$ distinguishes $i'w'$ and $j'w'$ with respect to $S'$, we see that $i'w'x' \in S \iff j'w'x' \not\in S$.
\item
Hence $(s,i'w'x') \in S \times S' \iff (s,j'w'x') \in S \times \ol{S'}$.
\ei
It follows that either $w$ or $wx$ distinguishes $(i,i')$ and $(j,j')$ with respect to $X$, regardless of whether we have $X = S \times S'$ or $X = (S \times S') \cup (\ol{S} \times \ol{S'})$.

\bm{Case 2 (States in the same column)}:
Suppose $i' = j'$, that is, the states are in the same column but different rows.
Since $G$ is primitive, all permutation DFAs with state set $Q$ and transition group $G$ are uniformly minimal.
Hence there exists $x \in G$ that distinguishes $i$ and $j$ with respect to $S$.
Suppose without loss of generality that $ix \in S$ and $jx \not\in S$.
Since $R_{ix,jx}\pi'$ is primitive, it is transitive, and thus there exists $y' \in R_{ix,jx}\pi'$ such that $i'x'y' \in S$.
If $y$ fixes $ix$ and $jx$,
then $(ixy,i'x'y') = (ix,i'x'y') \in S \times S'$, and $(jxy,j'x'y') = (jx,j'x'y') \in \ol{S} \times S'$ since $i' = j'$ implies $i'x'y' = j'x'y'$.
If $y$ swaps $ix$ and $jx$
then $(ixy,i'x'y') = (jx,i'x'y') \in \ol{S} \times S'$ and $(jxy,j'x'y') = (ix,j'x'y') \in S \times S'$.
In either case, it follows that $xy$ distinguishes $(i,i')$ and $(j,j')$ with respect to $X$, regardless of whether we have $X = S \times S'$ or $X = (S \times S') \cup (\ol{S} \times \ol{S'})$.

\bm{Case 3 (States in different rows and different columns)}:
Suppose $i \ne i'$ and $j \ne j'$.
We divide this case into two subcases.

\bm{Subcase 3a ($X = S \times S'$)}:
We may assume without loss of generality that $(i,i')$ and $(j,j')$ are both in $X$.
To see this, observe that since $G$ is transitive and $R_{q}\pi' \le G'$ is transitive for each $q \in Q$, we know that $G_\times$ is transitive by Lemma \ref{lem:trans}.
Thus there exists $w \in \Sig^*$ that maps $(i,i')$ to a state in $X$.
Hence either $w$ distinguishes the states, or $w$ also maps $(j,j')$ into $X$.

Suppose we have $(i,i'),(j,j') \in X = S \times S'$.
Since $R_{i,j}\pi'$ is primitive, there exists $w \in R_{i,j}\pi'$ that distinguishes $i'$ and $j'$ with respect to $S'$.
Without loss of generality, assume $i' \in S'$ and $j' \not\in S'$.
Then $(i,i')(w,w') = (i,i') \in S \times S'$ and $(j,j')(w,w') \in S \times \ol{S'}$.
Hence $w$ distinguishes the states.

\bm{Subcase 3b ($X = (S \times S') \cup (\ol{S} \times \ol{S'})$)}:
This is the final case we must deal with, and most complicated part of the proof.
We introduce a notion of \e{polarity} to simplify the arguments.
We assign a polarity of $1$ or $-1$ to each state in $Q \times Q'$ as follows.
First, let $q \in Q$ have polarity 1 if $q \in S$ and polarity $-1$ if $q \not\in S$.
Similarly, $q' \in Q'$ has polarity 1 if $q' \in S'$ and polarity $-1$ if $q' \not\in S'$.
Then the polarity of $(q,q') \in Q \times Q'$ is the product of the polarities of $q$ and $q'$.

Next, we partition $Q \times Q'$ into four \e{quadrants}: $S \times S'$, $S \times \ol{S'}$, $\ol{S} \times S'$, and $\ol{S} \times \ol{S'}$. Notice that in each quadrant, all states have the same polarity. Furthermore, the set $X = (S \times S') \cup (\ol{S} \times \ol{S'})$ is the set of all states with \e{positive} polarity, and the set $\ol{X} = (S \times \ol{S'}) \cup (\ol{S} \times S')$ is the set of all states with \e{negative} polarity. Hence to show all states are distinguishable by $X$, we must show that for each pair of states of equal polarity, there is a word that preserves the polarity of one state and reverses the polarity of the other state.

We now prove two claims, which together complete the proof of this subcase. First we show that pairs of states in the same quadrant are distinguishable, and then we show that pairs of states in different quadrants are distinguishable.

\bm{Claim 1 (States in the same quadrant are distinguishable)}:
Suppose $(i,i')$ and $(j,j')$ are in the same quadrant. 
This means that $(i,i')$ and $(j,j')$ have the same polarity; furthermore, $i$ and $j$ have the same polarity, and $i'$ and $j'$ have the same polarity. 
\bi
\item
Choose a word $w' \in R_{i,j}\pi'$ that distinguishes $i'$ and $j'$ with respect to $S'$ (by primitivity of $R_{i,j}\pi'$).
\item
Notice that $w$ preserves the polarity of both $i$ and $j$, since it either fixes both $i$ and $j$ or it swaps them, and $i$ and $j$ have the same polarity.
\item
Since $(i,i')$ and $(j,j')$ are in the same quadrant, we either have $i',j' \in S'$ or $i',j' \in \ol{S'}$.
\item
Since $w'$ distinguishes $i'$ and $j'$ with respect to $S'$, it follows that $w'$ acts on $i'$ and $j'$ by preserving the polarity of one state and reversing the polarity of the other.
\ei
It follows that $(w,w')$ acts on $(i,i')$ and $(j,j')$ by preserving the polarity of one state and reversing the polarity of the other.
In other words, $w$ distinguishes these states.


\bm{Claim 2 (States in different quadrants are distinguishable)}:
Suppose that $(i,i')$ and $(j,j')$ lie in different quadrants. 
\bi
\item
We may assume without loss of generality that $(i,i')$ and $(j,j')$ have the same polarity; otherwise they are trivially distinguishable.
\item
We may also assume without loss of generality that $(i,i'),(j,j') \in X$, by the same argument we used in Subcase 3a.
Thus we must have $(i,i') \in S \times S'$ and $(j,j') \in \ol{S} \times \ol{S'}$, or vice versa.
\item
We may assume without loss of generality that $(i,i') \in S \times S'$ and $(j,j') \in \ol{S} \times \ol{S'}$, by swapping the names of $(i,i')$ and $(j,j')$ if necessary.
\ei
So we have reduced to the case where one state is in quadrant $S \times S'$ and the other is in quadrant $\ol{S} \times \ol{S'}$.

\bi
\item
Since $G$ is primitive, $S$ and $\ol{S}$ are either not blocks, or they are trivial blocks. 
\item
$S$ and $\ol{S}$ are proper non-empty subsets of $Q$, so they can only be trivial blocks if $|S| = |\ol{S}| = 1$.
\item
This would imply $|Q| = |S| + |\ol{S}| = 2$, and we are assuming $|Q| \ge 3$, so they cannot both be trivial blocks. So at least one of $S$ or $\ol{S}$ is not a block.
Note that in the case where $G'$ is primitive and the groups $C_{i',j'}\pi \le G$ are primitive, we would use $|Q'| \ge 3$ here.
\ei
If $S$ is not a block, let $w \in G$ be a word such that $\emp \subsetneq Sw \cap S \subsetneq S$.
Otherwise, $\ol{S}$ is not a block, so let $w \in G$ be a word such that $\emp \subsetneq \ol{S}w \cap \ol{S} \subsetneq \ol{S}$.

Now, we partition $Q$ into two sets:
\[ 
P = \{ q \in S : qw \in S \} \cup \{q \in \ol{S} : qw \in \ol{S}\},\]
\[
\ol{P} = \{ q \in S : qw \in \ol{S} \} \cup \{q \in \ol{S} : qw \in S\}.
\]
Note that $P$ is non-empty, since if it was empty we would have $Sw \cap S = \emp$ and $\ol{S}w \cap \ol{S} = \emp$.
Similarly, $\ol{P}$ is non-empty, since otherwise we would have $Sw \cap S = S$ and $\ol{S}w \cap \ol{S} = \ol{S}$.

Observe that if $i,j \in P$, then $P$ is a proper subset of $Q$ with size at least two, so it cannot be a block for $G$. Hence some word in $G$ distinguishes $i$ and $j$ by $P$.
Similarly, if $i,j \in \ol{P}$, then $i$ and $j$ are distinguishable by $\ol{P}$.
So we may assume that either $i \in P$ and $j \in \ol{P}$, or $i \in \ol{P}$ and $j \in P$.

Suppose that $i \in P$ and $j \in \ol{P}$.
Recall that we have $(i,i') \in S \times S'$ and $(j,j') \in \ol{S} \times \ol{S'}$.
Thus since $i \in S$ we have $iw \in S$, and since $j \in \ol{S}$ we have $jw \in S$.
Hence $(iw,i'w'),(jw,j'w') \in S \times Q'$, that is, $w$ maps both $(i,i')$ and $(j,j')$ into $S \times Q'$.
There are two possibilities: 
\bi
\item
The states $(iw,i'w')$ and $(jw,j'w')$ are in the same quadrant, and thus are distinguishable by Claim 1.
\item
The states $(iw,i'w')$ and $(jw,j'w')$ are in different quadrants.
Since $iw$ and $jw$ are both in $S$, one state must lie in $S \times S'$ and the other $S \times \ol{S'}$.
Thus $w$ distinguishes $(i,i')$ and $(j,j')$, and we are done.
\ei
So if $i \in P$ and $j \in \ol{P}$, we have proved the claim.
If we have $i \in \ol{P}$ and $j \in P$, then we have $iw \in \ol{S}$ and $jw \in \ol{S}$, and so a symmetric argument shows that the states are distinguishable.
This completes the proof of Claim 2.
By Claim 1 and Claim 2, 
we see that 
all pairs of states are distinguishable by sets of the form $(S \times S') \cup (\ol{S} \times \ol{S'})$, completing the proof of 
Subcase 3b and hence Case 3.

We have shown that for every pair of sets $\emp \subsetneq S \subsetneq Q$ and $\emp \subsetneq S' \subsetneq Q'$ and each $(S,S')$-compatible subset $X \subseteq Q \times Q'$, each pair of states in $(\cA \times \cA')(X)$ is distinguishable by $X$.
Thus $\cA \times \cA'$ is uniformly boolean minimal. \qed
\epf

While Lemma \ref{lem:boolprim} gives sufficient conditions for uniform boolean minimality, the conditions are not necessary. We will demonstrate this later, in Example \ref{ex:affine-bp}.

In the above proof, most of the difficulty came from dealing with the case $(S \times S') \cup (\ol{S} \times \ol{S'})$, which corresponds to the operation of symmetric difference (or complement of symmetric difference). In fact, if we choose to ignore the operation of symmetric difference, we can obtain necessary and sufficient conditions for the corresponding weaker version of uniform boolean minimality.

\bp
\label{prop:ns}
Suppose $\cA$ and $\cA'$ are permutation DFAs. The following are equivalent:
\be
\item
\label{ns:prim}
$R_{q}\pi' \le G'$ and $C_{q'}\pi \le G$ are primitive for all $q \in Q$ and $q' \in Q'$.
\item
\label{ns:min}
For all sets $\emp \subsetneq S \subsetneq Q$ and $\emp \subsetneq S' \subsetneq Q'$, the DFA $(\cA \times \cA')(S \times S')$ is minimal.
\ee
\ep

\bpf
\bm{$(\ref{ns:prim}) \implies (\ref{ns:min})$}: 
We proceed as in the proof of Lemma \ref{lem:boolprim}.
Let $(i,i')$ and $(j,j')$ be distinct states of $\cA \times \cA'$; we will show they are distinguishable with respect to $S \times S'$.

\bm{Case 1 (States in the same row)}:
In the proof of Lemma \ref{lem:boolprim}, we proved that states in the same row are distinguishable, using the facts that $G$ is transitive and $R_{q}\pi'$ is primitive for all $q \in Q$. 
Those facts still hold under our new hypotheses, so the same argument can be used here.

\bm{Case 2 (States in the same column)}:
The argument we used for this case in Lemma \ref{lem:boolprim} relied on the double row stabilizers being transitive, so we cannot use it here.
However, under our new hypotheses, we know that $G'$ is transitive and $C_{q'}\pi$ is primitive for all $q' \in Q$.
Thus we can just use a symmetric argument to the one for states in the same row.

\bm{Case 3 (States in different rows and different columns)}:
Suppose $i \ne j$ and $i' \ne j'$.
As in the proof of Lemma \ref{lem:boolprim}, we may assume without loss of generality that $(i,i'),(j,j') \in S \times S'$.
Since $C_{i'}\pi'$ is primitive, there exists $w \in C_{i'}\pi$ that distinguishes $i$ and $j$ with respect to $S$.
We claim that we can choose $w$ so that $iw \in S$ and $jw \not\in S$:
\bi
\item
Suppose for a contradiction that for all words $w \in C_{i'}\pi$ which distinguish $i$ and $j$ by $S$, we have $iw \not\in S$ and $jw \in S$.
\item
Fix $x \in C_{i'}\pi$ such that $ix \not\in S$ and $jx \in S$. 
\item
By transitivity of $C_{i'}\pi$, we can choose $y \in C_{i'}\pi$ such that $(ixy,i') \tr{y} (jx,i')$. 
It follows $ixy = jx$, and thus $ixy \in S$.
\item
Suppose $jxy \not\in S$. Then we have $ixy \in S$ and $jxy \not\in S$, where $xy \in C_{i'}\pi$. This contradicts our assumption that for all words $w \in C_{i'}\pi$ which distinguish $i$ and $j$ by $S$, we must have $iw \not\in S$ and $jw \in S$.
\item
Thus we can assume $jxy \in S$.
But then since $ixy = jx$, we have $ixy^2 = jxy \in S$. If $jxy^2 \not\in S$, we get a contradiction as before, so $jxy^2 \in S$.
\item
In general, we have $ixy^k = jxy^{k-1}$, and it follows by induction on $k$ that $ixy^k \in S$ and $jxy^k \in S$ for all $k > 1$.
\item
But $C_{i'}\pi$ is a finite group, so $y^k = \eps$ for some $k$. This gives $ixy^k = ix \in S$ and $jxy^k = jx \in S$, which contradicts the fact that $ix \not\in S$.
\ei
This proves the claim; there must exist $w \in C_{i'}\pi$ such that $iw \in S$ and $jw \not\in S$.
Then we have $(i,i') \tr{w} (iw,i') \in S \times S'$, but $(j,j') \tr{w} (jw,j'w') \not\in S \times S'$ since $jw \not\in S$.
Thus $w$ distinguishes the states.
We have now shown that all pairs of states $(i,i')$ and $(j,j')$ are distinguishable.

\bm{$(\ref{ns:min}) \implies (\ref{ns:prim})$}: 
Suppose that for all sets $\emp \subsetneq S \subsetneq Q$ and $\emp \subsetneq S' \subsetneq Q'$, the DFA $(\cA \times \cA')(S \times S')$ is minimal.
Assume for a contradiction that there exists $q' \in Q'$ such that $C_{q'}\pi$ is not primitive.
Then there exists a non-trivial block $B \subsetneq Q$ for $C_{q'}\pi$.
We claim that $(\cA \times \cA)(B \times \{q'\})$ is not minimal:
\bi
\item
Since $B$ is a non-trivial block, it contains at least two distinct elements.
Let $i,j \in B$ be distinct and consider the states $(i,q')$ and $(j,q')$ of $\cA \times \cA'$.
\item
If $q'w' \ne q'$, then $w$ does not distinguish $(i,q')$ and $(j,q')$. Indeed, if $q'w' \ne q'$, then $(iw,q'w')$ and $(jw,q'w')$ both lie outside of $B \times \{q'\}$.
\item
Hence if $w \in \Sig^*$ distinguishes $(i,q')$ and $(j,q')$, we must have $q'w' = q'$, and thus $w \in C_{q'}\pi$.
\item
Since $B$ is a block for $C_{q'}\pi$, we either have $iw,jw \in B$ (if $Bw = B$) or $\{iw,jw\} \cap B = \emp$ (if $Bw \cap B = \emp$).
\item
Thus $w \in C_{q'}\pi$ cannot distinguish $(i,q')$ and $(j,q')$: if $iw,jw \in B$ then $w$ maps both states to $B \times \{q\}$; otherwise it maps both states to $\ol{B} \times \{q\}$.
\item
It follows that no word can distinguish $(i,q')$ and $(j,q')$ by $B \times \{q'\}$.
\ei
This shows that $(\cA \times \cA')(B \times \{q'\})$ is not minimal, which is a contradiction.
It follows that $C_{q'}\pi$ must be primitive for all $q' \in Q'$.
A symmetric argument shows that $R_{q}\pi'$ must be primitive for all $q \in Q$. \qed
\epf

One may wonder whether condition (\ref{ns:prim}) of Proposition \ref{prop:ns} are actually sufficient to prove Lemma \ref{lem:boolprim}.
We will show later (Example \ref{ex:affine-ns}) that this is not the case.

\subsection{Dissimilar DFAs}
\label{sec:dissimilar}
While the conditions of Lemma \ref{lem:boolprim} are somewhat complicated, there are cases where we can easily verify that they hold. We consider some of these cases next.

We will say that the DFAs $\cA$ and $\cA'$ are \e{similar} if the maps $\pi \co M_\times \ra M$ and $\pi' \co M_\times \ra M'$ are injective.
By Proposition \ref{prop:monoid}, $\pi$ and $\pi'$ are always surjective, so if they are also injective then they are isomorphisms. 
Hence if $\cA$ and $\cA'$ are similar, the map $\pi^{-1}\pi' \co M \ra M'$ given by $w \mapsto w'$ is a well-defined monoid isomorphism.
If $\cA$ and $\cA'$ are not similar, we say they are \e{dissimilar}.
If $\pi$ and $\pi'$ \e{both} fail to be injective, we say that $\cA$ and $\cA'$ are \e{strongly dissimilar}.

We give some examples of dissimilar DFAs.
All DFAs will be over the two-letter alphabet $\{a,b\}$, and we will not specify the initial and final states since they do not affect whether DFAs are similar.
\bx
Let $\cA$ have states $\{1,2\}$ and transformations $a = (1,2)$ and $b = ()$.
Let $\cA'$ have states $\{1,2\}$ and transformations $a' = ()$ and $b' = (1,2)$.
Notice that in the transition group of $\cA \times \cA'$, we have $(b,b') = ((),(1,2))$ and $(\eps,\eps') = ((),())$.
Thus $(b,b')\pi = (\eps,\eps')\pi = ()$ and it follows that $\pi$ is not injective.
Hence $\cA$ and $\cA'$ are dissimilar.
In fact, a symmetric argument shows that $\pi'$ is not injective, and thus these DFAs are strongly dissimilar.

Another way to see that these DFAs are dissimilar is to consider the ``map'' $w \mapsto w'$. 
This ``map'' is not actually well-defined, since from the fact that $b \mapsto b'$ we must have $() \mapsto (1,2)$, but from the fact that $\eps \mapsto \eps'$ we must have $() \mapsto ()$.
(Formally, this ``map'' is a \e{binary relation}; we say it is ``well-defined'' if the relation happens to be a function, and otherwise is not.)
This means $\cA$ and $\cA'$ cannot be similar, since we know that if they are similar, then $\pi^{-1}\pi'$ is a well-defined isomorphism which sends $w$ to $w'$.

Alternatively, without even checking whether the ``map'' $w \mapsto w'$ is well defined, we can see that since $() = b \mapsto b' = (1,2)$, this ``map'' sends an element of order one to an element of order two, and thus it cannot possibly be a group isomorphism. But then $\cA$ and $\cA'$ cannot be similar.
\qedb
\ex

Usually, the easiest way to prove that a pair of DFAs is dissimilar is to examine the ``map'' $w \mapsto w'$ and show that either it is not well-defined or not an isomorphism. 

\bx
Let $\cA$ have states $\{1,2\}$ and transformations \mbox{$a = (1,2)$, $b = (1,2)$.}
Let $\cA'$ have states $\{1,2,3,4\}$ and transformations $a'\,{=}\,(1,2)(3,4)$, $b\,{=}\,(1,3)(2,4)$.
The transition group of $\cA$ has two elements: $\eps = ()$ and $a = b = (1,2)$.
However, the transition group of $\cA'$ has four elements: 
\[ \eps = (),\;a' = (1,2)(3,4),\;b' = (1,3)(2,4),\;a'b' = (1,4)(2,3). \]
Hence these DFAs must be dissimilar, since they have different transition groups.
Similar DFAs always have isomorphic transition monoids/groups.

These DFAs are not strongly dissimilar.
To see this, observe that the transition group of $\cA \times \cA'$ has four elements:
\[ 
(\eps,\eps') = ((),()),\quad
(a,a') = ((1,2),(1,2)(3,4)), \]
\[ 
(b,b') = ((1,2),(1,3)(2,4)),\quad
(ab,a'b') = ((),(1,4)(2,3)). 
 \]
It is easy to verify that any other product of elements will be equal to one of these four. Now note that $\pi'$ is a bijection but $\pi$ is not, and thus $\cA$ and $\cA'$ are not strongly dissimilar.
In this case, the ``map'' $w \mapsto w'$ is not well-defined.
However, the map $(\pi')^{-1}\pi$ given by $w' \mapsto w$ \e{is} well-defined, and in fact is a group homomorphism (but not an isomorphism). \qedb
\ex

As for examples of similar DFAs, we have the following fact: isomorphic DFAs are necessarily similar. Indeed, suppose there is an isomorphism $f \co Q \ra Q'$. Then for all $q \in Q$ and $a \in \Sig$, we have $(qa)f = (qf)a'$, and thus $qa = q(fa'f^{-1})$.
It follows that $a = fa'f^{-1}$ for all $a \in \Sig^*$, and thus $w = fw'f^{-1}$ for all $w \in \Sig^*$.
Hence $\pi \co M_\times \ra M$ is given by $(w,w')\pi = (fw'f^{-1},w')\pi = fw'f^{-1}$, and this map is clearly injective since if $fw'f^{-1} = fx'f^{-1}$ then $w' = x'$.
Similarly, we have $w' = f^{-1}wf$ and $\pi' \co M_{\times} \ra M$ is injective.

This shows that DFA similarity is a generalization of DFA isomorphism. However, the next example shows that it is possible for two DFAs with different numbers of states to be similar, and that the monoid isomorphism $w \mapsto w'$ does not necessarily have to be conjugation by a permutation.

\bx
\label{ex:similar-it}
Consider the symmetric group $S_{10}$. This group contains an intransitive subgroup that is isomorphic to $S_{5}$, given by permutations of $\{1,\dotsc,10\}$ which fix every point in $\{6,\dotsc,10\}$.
This can be considered the ``natural'' embedding of $S_5$ in $S_{10}$.
However, there is also a primitive subgroup of $S_{10}$ that is isomorphic to $S_5$, and is not conjugate to this natural embedding.
In GAP's library of primitive groups, this subgroup can be accessed with the command {\tt PrimitiveGroup(10,2)}.
We can use GAP to compute an explicit isomorphism between $S_5$ and this subgroup:
\begin{verbatim}gap> IsomorphismGroups(SymmetricGroup(5),PrimitiveGroup(10,2));
[ (3,4), (1,2,3)(4,5) ] -> 
[ (1,2)(6,8)(7,9), (2,6,4,5,3,7)(8,10,9) ]
\end{verbatim}
The GAP output tells us that the map which sends $(3,4)$ to $(1,2)(6,8)(7,9)$ and $(1,2,3)(4,5)$ to $(2,6,4,5,3,7)(8,10,9)$ can be extended multiplicatively to an isomorphism between $S_5$ and the aforementioned primitive subgroup of $S_{10}$.
We use this isomorphism to construct similar permutation DFAs of different sizes.

Let $\cA$ have states $\{1,\dotsc,5\}$ and transformations $a = (3,4)$, $b = (1,2,3)(4,5)$.
Let $\cA'$ have states $\{1,\dotsc,10\}$ and transformations $a' = (1,2)(6,8)(7,9)$ and $b' = (2,6,4,5,3,7)(8,10,9)$.

The transition group $G$ of $\cA$ is $S_5$, and the transition group $G'$ of $\cA'$ is a primitive subgroup of $S_{10}$ that is isomorphic to $S_5$.
Furthermore, the map $w \mapsto w'$ is an isomorphism of $G$ and $G'$.
Hence $\cA$ and $\cA'$ are similar.
This pair of DFAs has other interesting properties; we will revisit them in Example \ref{ex:similar-it2}.

Notice that similarity of DFAs is a very fragile property;
if we simply switch the roles of $a'$ and $b'$ in $\cA'$, giving $a = (3,4)$ and $a' = (2,6,4,5,3,7)(8,10,9)$, then $\cA$ and $\cA'$ are no longer similar.
Indeed, after the switch, wee see that $w \mapsto w'$ sends an element of order two to an element of order six, which means it cannot be an isomorphism of $G$ and $G'$. \qedb
\ex

Dissimilar permutation DFAs have the following nice property.

\bp
\label{prop:dissimilar}
If $\cA$ and $\cA'$ are dissimilar permutation DFAs, then at least one of the following statements holds: 
\be
\item
$R\pi'$ is a non-trivial normal subgroup of $G'$.
\item
$C\pi$ is a non-trivial normal subgroup of $G$.
\ee
If $\cA$ and $\cA'$ are strongly dissimilar, then both hold.
\ep
\bpf
Recall that $R = \ker\pi$ and $C = \ker\pi'$; these are normal subgroups of $G_\times$. Since $\pi$ and $\pi'$ are surjective, $R\pi'$ is a normal subgroup of $G'$ and $C\pi$ is a normal subgroup of $G$.
\bi
\item
We have $\ker\pi = \{ (w,w') \in G_\times : w = \eps\}$, so $R\pi' = \{ w' \in G' : w = \eps \}$ and similarly $C\pi = \{w \in G : w' = \eps'\}$.
\item
If $R\pi'$ is trivial, then whenever $w = \eps$ we have $w' = \eps'$, and so $R = \ker\pi = \{(\eps,\eps')\}$ is trivial; hence $\pi$ is injective.
\item
Similarly, if $C\pi$ is trivial, then $C = \ker\pi'$ is trivial, and thus $\pi'$ is injective.
\ei
Thus we see that if $\cA$ and $\cA'$ are dissimilar, then $\pi$ and $\pi'$ cannot both be injective, and so $R\pi'$ and $C\pi$ cannot both be trivial.
Furthermore, if $\cA$ and $\cA'$ are strongly dissimilar, then neither $\pi$ nor $\pi'$ can be injective, and so neither $R\pi'$ nor $C\pi$ can be trivial.
\qed
\epf

This leads to a useful theorem:
\bt
\label{thm:dissimilar}
Let $\cA$ and $\cA'$ be dissimilar permutation DFAs with $|Q|,|Q'|\ge 3$.
\be
\item
\label{di:trans}
Suppose $G$ and $G'$ are transitive.
If all non-trivial normal subgroups of $G$ and of $G'$ are transitive, then $\cA \times \cA'$ is accessible.
\item
\label{di:prim}
Suppose $G$ and $G'$ are primitive.
If all non-trivial normal subgroups of $G$ and of $G'$ are primitive, then $\cA \times \cA'$ is uniformly Boolean minimal.
\ee
\et
\bpf
By Proposition \ref{prop:dissimilar}, since $\cA$ and $\cA'$ are dissimilar, one of $R\pi'$ or $C\pi$ is a non-trivial normal subgroup.
Suppose that $C\pi \le G$ is non-trivial; the other case is symmetric.

\bm{(\ref{di:trans})}: Since all non-trivial normal subgroups of $G$ are transitive, $C\pi$ is transitive. 
Hence $C_{q'}\pi$ is transitive for all $q' \in Q$, since $C_{q'}\pi \ge C\pi$.
Since $G'$ is transitive, we see that condition (\ref{lt:rowcol}) of Lemma \ref{lem:trans} holds.
Thus $\cA \times \cA'$ is accessible.

\bm{(\ref{di:prim})}: Since all non-trivial normal subgroups of $G$ are primitive, $C\pi$ is primitive.
Hence $C_{p',q'}\pi$ is primitive for all $p',q' \in Q$, since $C_{p',q'}\pi \ge C\pi$.
Since $G'$ is primitive, we see that $\cA$ and $\cA'$ meet the conditions of Lemma \ref{lem:boolprim}.
Thus $\cA \times \cA'$ is uniformly Boolean minimal.
\qed
\epf

The power of Theorem \ref{thm:dissimilar} comes from the fact that several interesting classes of groups have the property that all non-trivial normal subgroups are transitive or primitive.
The next corollary gives examples of when Theorem \ref{thm:dissimilar} can be applied.

\bc
\label{cor:dissimilar}
Let $\cA$ and $\cA'$ be dissimilar permutation DFAs with $|Q|,|Q'| \ge 3$.
\be
\item
Suppose $G$ and $G'$ are transitive.
\be
\item
\label{cd:trs}
If $G$ and $G'$ are transitive simple groups, then $\cA \times \cA'$ is accessible.
\item
\label{cd:trp}
If $G$ and $G'$ are primitive groups, then $\cA \times \cA'$ is accessible.
\ee
\item
Suppose $G$ and $G'$ are primitive.
\be
\item
\label{cd:prs}
If $G$ and $G'$ are primitive simple groups, then $\cA \times \cA'$ is uniformly Boolean minimal.
\item
\label{cd:psa}
If $G$ is $S_Q$ or $A_Q$, and $G'$ is $S_{Q'}$ or $A_{Q'}$, and we have $(|Q|,|Q'|) \not\in \{(3,4),(4,3),(4,4)\}$, then $\cA \times \cA'$ is uniformly Boolean minimal.
\item
\label{cd:p2t}
If $G$ and $G'$ are 2-transitive groups which are not of affine type, then $\cA \times \cA'$ is uniformly Boolean minimal. 
\ee
\ee
\ec
Before proving this, we will explain what we mean by ``affine type''. The notion of ``affine type'' comes from the O'Nan-Scott theorem~\cite[Theorem 4.1A]{DiMo96}, a structure theorem for primitive groups. The O'Nan-Scott theorem divides the primitive groups into different types based on their \e{socle}.
The socle of a group $G$ is the subgroup generated by all the \e{minimal normal subgroups} of $G$, that is, the normal subgroups $N$ of $G$ for which there does not exist a non-trivial normal subgroup $N'$ of $G$ with $N' \subseteq N$.

A group is \e{abelian} if its binary operation is commutative.
A primitive group with an abelian socle is necessarily of \e{affine type}, which means it is a permutation group of degree $p^d$ for $p$ prime and $d \ge 1$, and is isomorphic to a subgroup of the \e{affine group} $AGL(d,p)$.
This is a group of permutations of the set $\bF^d_p$, where $\bF_p$ is the finite field with $p$ elements; it consists of all maps of the form $(\gamma_1,\dotsc,\gamma_d) \mapsto (\alpha \gamma_1 + \beta,\dotsc,\alpha \gamma_d + \beta)$, where $\alpha,\beta,\gamma_1,\dotsc,\gamma_d \in \bF_p$. 
After this proof, we will show that there exist DFAs $\cA$ and $\cA'$ whose transition groups are 2-transitive groups of affine type, such that $\cA \times \cA'$ is not uniformly Boolean minimal.
Hence
in the statement of Corollary \ref{cor:dissimilar} (\ref{cd:p2t}),
excluding 2-transitive groups of affine type is necessary.
We remark that $A_3$, $S_3$, $A_4$ and $S_4$ happen to be groups of affine type; we will see in the proof that this is not true for symmetric and alternating groups of larger degree.

\bpf
\bm{(\ref{cd:trs})}: Recall that a group is \e{simple} if it has no non-trivial proper normal subgroups. Hence if $G$ is a transitive simple group, then the only non-trivial normal subgroup of $G$ is $G$ itself. Similarly, the only non-trivial normal subgroup of $G'$ is $G'$ itself. Thus the conditions of Theorem \ref{thm:dissimilar} hold.

\bm{(\ref{cd:trp})} Suppose $G$ and $G'$ are primitive. 
It is an easy exercise in group theory to show that all non-trivial normal subgroups of a primitive group are transitive (e.g., see~\cite[Theorem 1.6A]{DiMo96}).
Thus the conditions of Theorem \ref{thm:dissimilar} hold in this case.

\bm{(\ref{cd:prs})}: 
Since $G$ is a primitive simple group, then the only non-trivial normal subgroup of $G$ is $G$ itself, and similarly for $G'$. Thus the conditions of Theorem \ref{thm:dissimilar} hold.

\bm{(\ref{cd:psa})}: First suppose $|Q| \ne 4$ and $|Q'| \ne 4$. 
It is well-known that $A_n$ is simple for $n \ne 4$ (e.g., see~\cite[Corollary 3.3A]{DiMo96}). Thus $A_Q$ and $A_{Q'}$ are primitive simple groups. So if $G = A_Q$ and $G' = A_{Q'}$, this case follows from (\ref{cd:prs}). 

Now, consider the symmetric groups.
It is well-known that for $n \ne 4$, the only non-trivial normal subgroups of $S_n$ are $A_n$ and $S_n$. 

To see this,
let $N$ be a non-trivial normal subgroup of $S_n$.
If $A_n \le N$, we claim that either $N = A_n$ or $N = S_n$.
Indeed, if $N$ properly contains $A_n$, then it contains some permutation $p$ that cannot be written as a product of an even number of 2-cycles. But every permutation can be written as a product of 2-cycles, so $p$ can be written as a product of an odd number of 2-cycles. Thus for each 2-cycle $(i,j)$, we have $(i,j)p \in A_n \le N$.
So $N$ contains $(i,j)pp^{-1} = (i,j)$; since $N$ contains all 2-cycles, we must have $N = S_n$.

If $N \le A_n$, consider $ghg^{-1}$ for $g \in A_n$ and $h \in N$.
Since $g$, $h$ and $g^{-1}$ are all in $A_n$, each can be written a product of an even number of 2-cycles, and thus $ghg^{-1}$ can be written this way as well.
So $N$ is also a non-trivial normal subgroup of $A_n$; but $A_n$ is simple for $n \ne 4$, so we must have $N = A_n$.

Thus a non-trivial normal subgroup is either $S_n$ or $A_n$. both of which are primitive.
Thus if $(G,G') \in \{(A_Q,S_{Q'}), (S_Q,A_{Q'}), (S_Q,S_{Q}')\}$, Theorem \ref{thm:dissimilar} applies and gives the result.
Note we have also shown that for $n \ge 5$, the alternating group $A_n$ is the unique minimal normal subgroup of $S_n$ and $A_n$. Thus the socle of $S_n$ and $A_n$ is non-abelian, which shows that $S_n$ and $A_n$ are not of affine type when $n \ge 5$.

Now, suppose $|Q| = 4$ or $|Q'| = 4$ and $(|Q|,|Q'|) \not\in \{(3,4),(4,3),(4,4)\}$. Then $|Q| \ge 5$ or $|Q'| \ge 5$.
Assume without loss of generality that $|Q| = 4$ and $|Q'| \ge 5$; the other case is symmetric.
Consider the normal subgroup $R\pi'$ of $G'$.
If $R\pi'$ is trivial, then as we argued in the proof of Proposition \ref{prop:dissimilar}, the map $\pi \co G_\times \ra G$ must be injective. Since $G$ is either $A_4$ or $S_4$, this means $|G_\times| \le |S_4| = 24$. But the map $\pi' \co G_\times \ra G'$ is surjective, and $G$ is either $A_{Q'}$ or $S_{Q'}$ for $|Q'| \ge 5$. This means $|G_\times| \ge |A_5| = 60$. So we have $60 \le 24$, which is a contradiction.
Thus $R\pi'$ cannot be trivial.
So $R\pi'$ is a non-trivial normal subgroup of $G'$, and thus it is primitive, since $G'$ is $S_{Q'}$ or $A_{Q'}$ and $|Q'| \ge 5$.
Thus the arguments in the proof of Theorem \ref{thm:dissimilar} apply and $\cA \times \cA'$ is uniformly Boolean minimal.

\bm{(\ref{cd:p2t})}: The results from permutation group theory that we use for this case are somewhat more advanced.
We will need the fact that a 2-transitive group has a unique minimal normal subgroup; this follows from two theorems in~\cite{DiMo96} (Theorem 4.1B and Theorem 4.3B), or alternatively is stated as a single result in~\cite{Cam81} (Proposition 5.2).
The other fact we need is that if the socle of a 2-transitive group is non-abelian, it is necessarily primitive~\cite[Theorem 7.2E]{DiMo96}. 

Now, suppose $G$ is 2-transitive.
The socle of $G$ is the subgroup generated by all the minimal normal subgroups; but $G$ has a unique minimal normal subgroup $N$, so the socle of $G$ is just equal to $N$.
If $N$ is abelian, then $G$ is of affine type.
Thus we may assume the socle $N$ is non-abelian; then it follows that $N$ is primitive.
Since $N$ is the unique minimal normal subgroup of $G$, every non-trivial normal subgroup of $G$ contains $N$, and thus is primitive.
Similarly, every non-trivial normal subgroup of $G'$ is primitive.
It follows that Theorem \ref{thm:dissimilar} applies.
\qed
\epf

\subsection{Affine Groups}
We now construct an infinite family of pairs of dissimilar permutation DFAs which have 2-transitive transition groups of affine type, and are not uniformly boolean minimal. The details of the construction require some knowledge of finite fields. We will first give the construction in full generality, and then use the construction to produce an explicit pair of 8-state DFAs.

\bx
\label{ex:affine}
For $k \ge 0$, let $\bF_{2^k}$ denote the finite field of order $2^k$.
For $\alpha,\beta,\xi \in \bF_{2^k}$ with $\alpha \ne 0$, define $t_{\alpha,\beta} \co \bF_{2^k} \ra \bF_{2^k}$ to be the map $\xi \mapsto \alpha\xi + \beta$.
The set of all such maps forms a group of permutations of $\bF_{2^k}$, which is called the \mbox{1-dimensional} \e{affine group} on $\bF_{2^k}$ and is denoted $AGL(1,2^k)$.
Multiplication (that is, composition of maps) in the affine group is given by the rule
\[ t_{\alpha,\beta}t_{\gamma,\xi} = t_{\alpha \gamma, \beta\gamma + \xi}. \]
It is an easy but somewhat tedious exercise to show that the affine group is 2-transitive. 
In~\cite[Chapter 4]{DiMo96} it is proved that the affine group has an abelian socle.

Recall that the multiplicative group of a finite field is cyclic.
Let $x$ be a generator for the multiplicative group of $\bF_{2^k}$.
We claim that the elements $t_{x,0}$ and $t_{1,1}$ generate $AGL(1,2^k)$.
This is once again an easy but tedious exercise, so we omit a proof.


An element of $AGL(1,2^k)$ of the form $t_{1,\beta}$ for some $\beta \in \bF_{2^k}$ is called a \e{translation}.
The translations form a subgroup of $AGL(1,2^k)$, which we call $T$.
We claim that the subgroup of translations $T$ is imprimitive and contains a block $B$ of size $2^{k-1}$.

To see this, recall that $\bF_{2^k}$ is a $k$-dimensional vector space over $\bF_2$.
Pick $k-1$ non-zero elements of $\bF_{2^k}$ and let $B$ be the subspace spanned by them.
Now consider $Bt_{1,\beta}$ for $\beta \in \bF_{2^k}$.
\bi
\item
If $\beta \in B$, then $\alpha+\beta \in B$ for all $\alpha \in B$, since $B$ is a subspace; thus $Bt_{1,\beta} = B$.
\item
If $\beta \not\in B$, then for all $\alpha \in B$ we have $\alpha+\beta \not\in B$.
Indeed, if $\alpha+\beta$ was in $B$ then $(\alpha+\beta)-\alpha = \beta$ would be in $B$, since $B$ is a subspace.
\item
Thus if $\beta \not\in B$, then $Bt_{1,\beta} \cap B = \emp$.
\ei
And so, we see that $B$ is indeed a block for $T$.


Consider the subgroup $A$ of $AGL(1,2^k) \times AGL(1,2^k)$ generated by the elements 
$a = (t_{x,0},t_{x,0})$, $b = (t_{1,1},t_{1,0})$ and $c = (t_{1,0},t_{1,1})$.
We claim that every element $(t_{\alpha,\beta},t_{\gamma,\xi})$ of $A$ has the property that $\alpha = \gamma$.
For simplicity, we will call elements with this property \e{balanced}.

Certainly the elements $a$, $b$ and $c$ are balanced, and so is the identity element $(t_{1,0},t_{1,0})$.
We will show that multiplying a balanced element on the right by $a$, $b$ or $c$ results in a balanced element.
Indeed, first observe that $t_{\alpha,\beta}t_{x,0} = t_{\alpha x,\beta x}$ and $t_{\alpha,\beta}t_{1,1} = t_{\alpha,\beta+1}$.
Thus if we take an arbitrary balanced element $(t_{\alpha,\beta},t_{\alpha,\gamma})$, then we have:
\[ 
(t_{\alpha,\beta},t_{\alpha,\gamma})a
=
(t_{\alpha,\beta}t_{x,0},t_{\alpha,\gamma}t_{x,0})
=
(t_{\alpha x,\beta x},t_{\alpha x,\gamma x}). \]
\[ 
(t_{\alpha,\beta},t_{\alpha,\gamma})b
=
(t_{\alpha,\beta}t_{1,1},t_{\alpha,\gamma}t_{1,0})
=
(t_{\alpha,\beta+1},t_{\alpha,\gamma}). \]
\[ 
(t_{\alpha,\beta},t_{\alpha,\gamma})c
=
(t_{\alpha,\beta}t_{1,0},t_{\alpha,\gamma}t_{1,1})
=
(t_{\alpha,\beta},t_{\alpha,\gamma+1}). \]
Since $a$, $b$ and $c$ and the identity are balanced, and multiplying a balanced element by $a$, $b$ or $c$ results in a balanced element, it follows the group $\gen{a,b,c} = A$ consists solely of balanced elements.


Next, let $\ol{B} = \bF_{2^k} \setminus B$ and consider the set $X = (B \times B) \cup (\ol{B} \times \ol{B}) \subseteq \bF_{2^k} \times \bF_{2^k}$. Notice that $(0,0),(1,1) \in \bF_{2^k} \times \bF_{2^k}$ both lie in $X$.
We claim that elements of $A$ cannot distinguish $(0,0)$ and $(1,1)$ with respect to $X$, that is, for all $g \in A$ we have $(0,0)g \in X \iff (1,1)g \in X$.
\bi
\item
To see this, consider an arbitrary element $g = (t_{\alpha,\beta},t_{\alpha,\gamma})$ of $A$.
\item
We have $0t_{\alpha,\beta} = \beta$ and $1t_{\alpha,\beta} = \alpha+\beta$. It follows that $(0,0)g = (\beta,\gamma)$ and $(1,1)g = (\alpha+\beta,\alpha+\gamma) = (\beta t_{1,\alpha},\gamma t_{1,\alpha})$.
\item
Since $B$ is a block for the subgroup of translations $T$, we either have $Bt_{1,\alpha} = B$ or $Bt_{1,\alpha} \cap B = \emp$.
\item
But $B$ has size $2^{k-1}$, which is exactly half the size of $\bF_{2^k}$, so if $Bt_{1,\alpha} \cap B = \emp$ then $Bt_{1,\alpha} = \ol{B}$.
\ei
It follows that if $(\beta,\gamma)$ is in $B \times B$ or $\ol{B} \times \ol{B}$ (that is, $(\beta,\gamma)$ is in $X$) then $(\beta t_{1,\alpha},\gamma t_{1,\alpha})$ is also in $B \times B$ or $\ol{B} \times \ol{B}$ (and thus in $X$).

Likewise, if $(\beta,\gamma)$ is not in $X$, then it is either in $B \times \ol{B}$ or $\ol{B} \times B$, and it follows $(\beta t_{1,\alpha},\gamma t_{1,\alpha})$ is not in $X$.
Thus $(0,0)g \in X \iff (1,1)g \in X$ for all $g \in A$.

Finally, for each $k \ge 1$, we construct a pair of DFAs over the alphabet $\{a,b,c\}$ with $2^k$ states each, which both have $AGL(1,2^k)$ as their transition group, but do not have a uniformly boolean minimal direct product.

We define a DFA $\cA$ as follows:
\bi
\item
The state set is $\bF_{2^k}$, the initial state is $0$, and the final state set is $B$.
\item
The transformations are $a = t_{x,0}$, $b = t_{1,1}$ and $c = t_{1,0}$.
\ei
We define $\cA'$ in the same way as $\cA$, except the roles of $b$ and $c$ are swapped:
\bi
\item
The transformations are $a' = t_{x,0}$, $b' = t_{1,0}$ and $c' = t_{1,1}$.
\ei
Since $t_{x,0}$ and $t_{1,1}$ generate $AGL(1,2^k)$, it is clear that both DFAs have $AGL(1,2^k)$ as their transition group.

Now consider $\cA \times \cA'$. Let $\circ$ be the ``complement of symmetric difference'' operation, so that $B \circ B = (B \times B) \cup (\ol{B} \times \ol{B})$. 
Observe that the transition group of $\cA \times \cA'$ is simply the group $A$.
Thus the states $(0,0)$ and $(1,1)$ of $\cA \times \cA'$ are not distinguishable by $B \circ B$. Hence $(\cA \times \cA')(B \circ B)$ is not minimal, and it follows that $\cA \times \cA'$ is not uniformly boolean minimal. \qedb
\ex

\bx
\label{ex:affine8}
We now carry out the construction of Example \ref{ex:affine} for $k=3$, $2^k = 8$.
First, we must construct the finite field $\bF_8$.
Let $\bF_2 = \{0,1\}$ denote the ring of integers modulo two.
We define $\bF_8$ to be the quotient ring $\bF_2[x]/\gen{x^3+x+1}$. 

This ring consists of polynomials of the form $\{ a+bx+cx^2 : a,b,c \in \bF_2\}$. Addition and multiplication work as usual for polynomials, except when multiplying, any terms of degree 3 or higher are reduced by repeatedly applying the rule $x^3 = x+1$.
Note also that since the coefficients of the polynomials are in $\bF_2$, we have $2f(x) = 0$ for all polynomials $f(x)$. For example, we have $(x^2+x+1) + (x^2+1) = x$.
Applying these facts, we see that:
\[ 
x^3 = x+1,\quad 
x^4 = x^2+x,\quad
x^5 = x^3+x^2= x^2+x+1,\]
\[
x^6 = x^3+x^2+x = x^2+x+x+1 = x^2+1,\quad
x^7 = x^3+x = x+1+x = 1. \]
These calculations show that every non-zero element of $\bF_8$ can be written as a power of the monomial $x$; in particular, $1$ can be written as a power of $x$, and thus $x$ is invertible. Hence $\bF_8$ is indeed a field and $x$ is a generator of  the multiplicative group of $\bF_8$.

Next, we explicitly write out the permutations $t_{x,0}$ and $t_{1,1}$ in cycle notation:
\[ t_{x,0} = (x,x^2,x^3,x^4,x^5,x^6,x^7) = (x,x^2,x+1,x^2+x,x^2+x+1,x^2+1,1). \]
\[ t_{1,1} = (0,1)(x,x+1)(x^2,x^2+1)(x^2+x,x^2+x+1) = (0,x^7)(x,x^3)(x^2,x^6)(x^4,x^5). \]
Finally, we need to find a block $B$ for the subgroup of translations of $AGL(1,8)$.
The construction tells us to pick two non-zero elements of $\bF_8$ and let $B$ be the subspace spanned by them.
If we take $1$ and $x$, we get $B = \{0,1,x,x+1\}$.

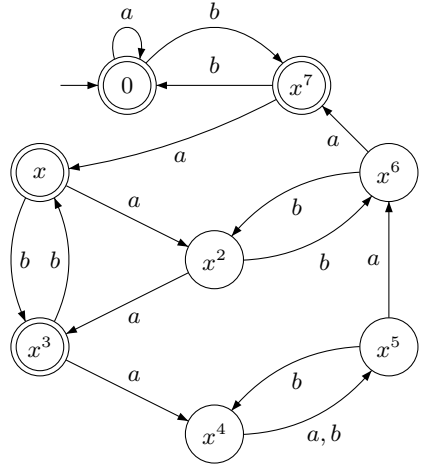
\begin{figure}[h]
\unitlength 11pt
\begin{center}\begin{picture}(12,13)(0,0)
\gasset{Nh=2,Nw=2,Nmr=1,ELdist=0.3,loopdiam=1}

\node(1)(3,12){$0$}\imark(1)
\node(2)(0,9){$x$}
\node(3)(6,6){$x^2$}
\node(4)(0,3){$x^3$}
\node(5)(6,0){$x^4$}
\node(6)(12,3){$x^5$}
\node(7)(12,9){$x^6$}
\node(8)(9,12){$x^7$}
\rmark(1)\rmark(2)\rmark(8)\rmark(4)

\drawedge(2,3){$a$}
\drawedge(3,4){$a$}
\drawedge(4,5){$a$}
\drawedge[curvedepth=-1, ELdist=-1.3](5,6){$a,b$}
\drawedge[curvedepth=-1](6,5){$b$}
\drawedge(6,7){$a$}
\drawedge(7,8){$a$}

\drawedge[ELdist=-1](8,1){$b$}
\drawedge[curvedepth=2](1,8){$b$}
\drawedge[curvedepth=0.5](8,2){$a$}

\drawedge[curvedepth=-1](2,4){$b$}
\drawedge[curvedepth=-1](4,2){$b$}
\drawedge[curvedepth=-1,ELdist=-1.1](3,7){$b$}
\drawedge[curvedepth=-1](7,3){$b$}

\drawloop(1){$a$}
\end{picture}\end{center}
\caption{DFA $\cA$ of Example~\ref{ex:affine8}. Each state also has a self-loop on letter $c$; these transitions are omitted from the diagram. The final state set is $B = \{0,x^7=1,x,x^3=x+1\}$, the block of the subgroup of translations of $AGL(1,8)$ that was found in Example \ref{ex:affine8}.}
\label{fig:affine-8}
\end{figure}

We now have all the information we need to construct $\cA$ and $\cA'$.
A state diagram for $\cA$ is shown in Figure \ref{fig:affine-8}, with the self-loops on $c$ omitted.
One may verify computationally that $\cA \times \cA'$ is not minimal when it is assigned the final state set $(B \times B) \cup (\ol{B} \times \ol{B})$. \qedb
\ex

For $k = 1$, we get DFAs $\cA$, $\cA'$ and $\cA \times \cA'$ that are isomorphic to the DFAs of Example \ref{ex:symdiff2}.
For $k = 2$, it happens that $AGL(1,4)$ is isomorphic to the alternating group $A_4$.
Hence the $k=2$ case gives an example of dissimilar DFAs that are not uniformly boolean minimal, and have alternating groups as their transition groups.
As Corollary \ref{cor:dissimilar} shows, this example does not generalize to alternating groups of higher degree.
The DFA $\cA$ of Example \ref{ex:altdfa} is isomorphic to the DFA $\cA$ produced by the construction of Example \ref{ex:affine} with $k = 2$.

The construction of Example \ref{ex:affine} also shows that condition (\ref{ns:prim}) of Proposition \ref{prop:ns} is not sufficient for uniform boolean minimality. 

\bx
\label{ex:affine-ns}
The DFAs $\cA$ and $\cA'$ constructed in Example \ref{ex:affine} are not uniformly boolean minimal.
However, we claim the subgroups $R_\alpha\pi' \le G'$ and $C_\alpha\pi \le G$ are primitive for all $\alpha \in \bF_{2^k}$, and thus $\cA$ and $\cA'$ meet condition (\ref{ns:prim}) of Proposition \ref{prop:ns}.
In fact, the groups $R_\alpha\pi'$ and $C_\alpha\pi$ are equal to $AGL(1,2^k)$.
First, we show that $R_0\pi'$ and $C_0\pi$ are equal to $AGL(1,2^k)$.
Consider $(a,a') = (t_{x,0},t_{x,0})$. 
\bi
\item
Since $t_{x,0}$ fixes $0$, it follows that $(a,a') \in R_0$ and $(a,a') \in C_0$.\\
Thus $a' = t_{x,0} \in R_0\pi'$ and $a = t_{x,0} \in C_0\pi$.
\item
Since $(b,b') = (t_{1,1},t_{1,0})$ and $b' = t_{1,0}$ fixes $0$, we see that $(b,b') \in C_0$.\\
Thus $b = t_{1,1} \in C_0\pi$.
\item
Similarly, since $(c,c') = (t_{1,0},t_{1,1})$, we see that $(c,c') \in R_0$.\\
Thus $c' = t_{1,1} \in R_0\pi'$.
\ei
Since $t_{x,0}$ and $t_{1,1}$ generate $AGL(1,2^k)$, and these elements are in $R_0\pi'$ and $C_0\pi$, it follows $R_0\pi'$ and $C_0\pi$ are equal to $AGL(1,2^k)$ and thus are primitive.

To show that $R_\alpha\pi'$ and $C_\alpha\pi$ are primitive for all $\alpha \ne 0$, we prove a general fact about single row and column stabilizers: if $C_{p'}\pi \le G$ is primitive for some $p' \in Q'$ and $G'$ is transitive, then $C_{q'}\pi$ is primitive for all $q' \in Q'$ (and similarly for single row stabilizers).
\bi
\item
To see this, choose $w' \in G'$ such that $p'w' = q'$.
\item
Let $B$ be a block for $C_{q'}\pi$.
We claim $Bw^{-1}$ is a block for $C_{p'}\pi$.
\item
To see this, choose $x \in C_{p'}\pi$ and consider
$Bw^{-1}x \cap Bw^{-1}$.
\item
If $Bw^{-1}x \cap Bw^{-1} \ne \emp$,
then $Bw^{-1}xw \cap B \ne \emp$.
\item
Now, for all $x \in C_{p'}\pi$, we have $p'x' = p'$ by definition.
\item
It follows $q' (w')^{-1}x'w' = p' x'w' = p'w' = q'$. Since $(w')^{-1}x'w'$ fixes $q'$, we have $w^{-1}xw \in C_{q'}\pi$.
\item
Since $B$ is a block for $C_{q'}\pi$ and $Bw^{-1}xw \cap B \ne \emp$,
we have $Bw^{-1}xw = B$.
\item
Hence $Bw^{-1}x = Bw^{-1}$, which proves $Bw^{-1}$ is a block for $C_{p'}\pi$.
\ei
It follows that if $B$ is a block for $C_{q'}\pi$, it must be a trivial block; otherwise $Bw^{-1}$ is a non-trivial block for the primitive group $C_{p'}\pi$, which is a contradiction.

Thus $C_\alpha\pi$ is primitive for all $\alpha \in \bF_{2^k}$, and symmetrically we see that $R_\alpha\pi'$ is primitive for all $\alpha \in \bF_{2^k}$.
This shows that $\cA$ and $\cA'$ satisfy condition (\ref{ns:prim}) of Proposition \ref{prop:ns}, yet $\cA \times \cA'$ is not uniformly boolean minimal. \qedb
\ex

We can also use DFAs derived from affine groups to show that the conditions of Lemma \ref{lem:boolprim} are not necessary for uniform boolean minimality.

\bx
\label{ex:affine-bp}
As in Example \ref{ex:affine8}, we define two DFAs $\cA$ and $\cA'$ that have transition group $AGL(1,8)$.
However, this time the direct product of the DFAs will be uniformly boolean minimal.
We define $\cA$ as follows (leaving the final state set unspecified).
\bi
\item
The state set is $\bF_{8}$, constructed as in Example \ref{ex:affine8}, and the initial state is $0$.
\item
The transformations are $a = t_{x,0}$ and $b = t_{1,1}$.
\ei
Define $\cA'$ to have the same states as $\cA$ and transformations $a' = t_{x,0}^{-1}$, $b' = t_{1,1}$.

Since $\cA$ and $\cA'$ only have 8 states each, we were able to verify computationally that
$\cA \times \cA'$ is uniformly boolean minimal by a brute force approach.
We computed $\cA \times \cA'$, and for each pair of sets $\emp \subsetneq S \subsetneq Q$ and $\emp \subsetneq S' \subsetneq Q'$, we checked the minimality of $(\cA \times \cA')(X)$ for all $(S,S')$-compatible sets $X$.

We have also verified computationally that $R_{0,1}\pi'$ and $C_{0,1}\pi$ are imprimitive, and hence $\cA$ and $\cA'$ do not meet the conditions of Lemma \ref{lem:boolprim}.
We verified this by using the cycle notation representation we found for $t_{x,0}$ and $t_{1,1}$ to explicitly construct the transition group $G_\times$ of $\cA \times \cA'$ in GAP.
Then we computed the setwise stabilizer $R_{0,1}$ of $\{0,1\} \times \bF_{8}$ (the rows indexed by $0$ and $1$), and the setwise stabilizer $C_{0,1}$ of $\bF_8 \times \{0,1\}$ (the columns indexed by $0$ and $1$).
Next, we computed $R_{0,1}\pi' \le G'$ and $C_{0,1}\pi \le G$.
These groups turned out to both be equal to $T$, the subgroup of translations in $AGL(1,8)$.
We saw earlier than $T$ is imprimitive. 
\qedb
\ex

We suspect that if this construction is generalized to $AGL(1,2^k)$, the resulting DFAs will have the same property of being uniformly boolean minimal but having $R_{0,1}\pi'$ and $C_{0,1}\pi$ imprimitive. However, we were unable to prove this.

\subsection{Similar DFAs}
To close out Section \ref{sec:main}, we consider what happens when the DFAs $\cA$ and $\cA'$ are similar. 
We have not investigated this case very deeply. In some ways, it seems much more difficult than the dissimilar case. Particularly, most of our results rely on the projections of various kinds of row and column stabilizers being transitive or primitive. For similar DFAs, the projections of the full row and column stabilizers $C\pi$ and $R\pi'$ are both trivial. Hence there is no guarantee that other types of stabilizers such as $R_q\pi'$ or $C_{p',q'}\pi$ have useful properties, or even that they are non-trivial, and so we cannot necessarily use these groups to our advantage.

On the other hand, similarity imposes the very strong condition that the groups $G$, $G'$ and $G_{\times}$ are all isomorphic. It may be possible to exploit this to prove some interesting things in the similar case.

It is not difficult to prove that if $\cA$ and $\cA'$ are isomorphic as DFAs, then $G_\times$ is necessarily intransitive, so the case of isomorphic similar DFAs is uninteresting for our purposes.
We give two examples demonstrating what can happen with non-isomorphic similar DFAs.

\bx
\label{ex:similar-it2}
Recall that in Example \ref{ex:similar-it}, we constructed two similar DFAs that are of different sizes (and hence are non-isomorphic) and have primitive transition groups.
\bi
\item
$\cA$ has states $\{1,\dotsc,5\}$ and transformations $a = (3,4)$, $b = (1,2,3)(4,5)$.
\item
$\cA'$ has states $\{1,\dotsc,10\}$ and transformations $a' = (1,2)(6,8)(7,9)$ and $b' = (2,6,4,5,3,7)(8,10,9)$.
\ei
We have verified computationally that $\cA \times \cA'$ has an intransitive transition group, and so is not accessible. 
Thus even if non-isomorphic similar DFAs have primitive transition groups, their direct product might have an intransitive transition group (compare this with Corollary \ref{cor:dissimilar} for dissimilar DFAs).

Note that in Example \ref{ex:similar-it}, we also showed that by simply swapping the roles of $a'$ and $b'$ in $\cA'$, the two DFAs $\cA$ and $\cA'$ become dissimilar. Furthermore, $\cA$ has the symmetric group $S_5$ as its transition group.
Thus by Corollary \ref{cor:dissimilar}, the two DFAs actually become uniformly boolean minimal if we swap the roles of $a'$ and $b'$. \qedb
\ex

\bx
\label{ex:similar}
There is a primitive subgroup of $S_6$ that is isomorphic to $S_5$. Using GAP, we can obtain an explicit isomorphism between $S_5$ and this subgroup, just as we did in Example \ref{ex:similar-it}.
\begin{verbatim}gap> IsomorphismGroups(SymmetricGroup(5),PrimitiveGroup(6,2));
[ (3,4), (1,2,3)(4,5) ] -> [ (1,2)(3,4)(5,6), (1,2,3,5,4,6) ]
\end{verbatim}
We then use this isomorphism to construct similar DFAs:
\bi
\item
$\cA$ has states $\{1,\dotsc,5\}$ and transformations $a = (3,4)$ and $b = (1,2,3)(4,5)$.
\item
$\cA'$ has states $\{1,\dotsc,6\}$ and transformations $a' = (1,2)(3,4)(5,6)$ and $b' = (1,2,3,4,5,6)$.
\ei
Unlike the DFAs of Example \ref{ex:similar-it2}, here we verified computationally that $\cA \times \cA'$ actually has a transitive transition group. Hence a direct product of non-isomorphic similar DFAs with primitive transition groups may or may not be accessible.

Note that $\cA \times \cA'$ is not uniformly boolean minimal. For example, we have verified computationally that $(\cA \times \cA')(X)$ is not minimal for $X = \{1\} \times \{1,3,5\}$.
We have not found an example of two similar DFAs that are uniformly boolean minimal, but we also have not proved that no such example exists. \qedb
\ex

\section{Conclusion}
\label{sec:conc}
We summarize the major results proved in Section \ref{sec:main}.

\bm{Theorem \ref{thm:1fstate}} gives necessary and sufficient conditions for a pair of regular languages $(L,L')$ to have maximal boolean complexity, with the requirement that these languages are recognized by permutation DFAs $\cA$ and $\cA'$ with exactly one final state. In this special case, it turns out that $(L,L')$ is uniformly boolean minimal if and only if $\cA \times \cA'$ is accessible. This gives a partial characterization of witnesses for the state complexity of proper binary boolean operations.

We have several results which may help to determine whether $\cA \times \cA'$ is accessible. \bm{Lemma \ref{lem:trans}} gives group-theoretic conditions for accessibility, while \bm{Proposition \ref{prop:graph}} gives a useful graph-theoretic condition.

\bm{Theorem \ref{thm:dissimilar}} gives a particularly useful group-theoretic condition for accessibility of $\cA \times \cA'$, as well as a similar condition for uniform boolean minimality. The power of this theorem is demonstrated by \bm{Corollary \ref{cor:dissimilar}}, which gives several classes of groups where the condition of Theorem \ref{thm:dissimilar} holds. If one can show that the transition group $\cA$ or of $\cA'$ lies in one of these classes, one immediately gets useful information about $\cA \times \cA'$. Corollary \ref{cor:dissimilar} is also useful for constructing examples of DFAs whose direct product is accessible or uniformly boolean minimal: one may pick a pair of groups from the classes mentioned in the corollary, and use them as the transition groups of a pair of DFAs.

\bm{Lemma \ref{lem:boolprim}} gives some sufficient conditions for uniform boolean minimality of permutation DFAs. The conditions  are stronger than those of Theorem \ref{thm:dissimilar}, but more difficult to verify. Unfortunately, Example \ref{ex:affine-bp} shows that these conditions are not necessary. Necessary and sufficient conditions for uniform boolean minimality are still unknown.

\bm{Proposition \ref{prop:ns}} gives a necessary and sufficient condition (\ref{ns:prim}) for a property that is slightly weaker than uniform boolean minimality to hold (in permutation DFAs).
Specifically, condition (\ref{ns:prim}) of Proposition \ref{prop:ns} does not guarantee minimality for final state sets corresponding to the symmetric difference operation or its complement.

Unfortunately, Example \ref{ex:affine-ns} shows that condition (\ref{ns:prim}) of Proposition \ref{prop:ns} is not sufficient for uniform boolean minimality.
This means a precise characterization of uniform boolean minimality lies strictly between the conditions given by Lemma \ref{lem:boolprim} and Proposition \ref{prop:ns}.

We state some unsolved problems and potential new directions of research arising from our work in this paper.
\bi
\item
Find necessary and sufficient conditions for $\cA \times \cA'$ to be uniform boolean minimal. Even in the special case of permutation DFAs, we were unable to resolve this.
\item
Find more classes of groups where the hypotheses of Theorem \ref{thm:dissimilar} hold, thus extending the reach of Corollary \ref{cor:dissimilar}.
\item
Find an example of two similar DFAs that are uniformly boolean minimal, or prove that no such DFAs exist. Even pairs of similar DFAs with an accessible direct product seem to be rare; one such pair is given in Example \ref{ex:similar}.
\item
Prove anything interesting about accessibility and/or uniform boolean minimality of direct products of non-isomorphic similar DFAs.
\item
Investigate uniform boolean minimality with respect to ``unrestricted state complexity'' (see the last two paragraphs of Section \ref{sec:def}).
\item
Investigate uniform boolean minimality for DFAs that do not contain ``interesting'' subgroups of permutations. 
That is, let us say the transition group of a DFA is ``interesting'' if it satisfies the hypotheses of any of our major results. Although we stated our results exclusively for permutation DFAs, they hold more generally for DFAs whose transition monoids contain an ``interesting'' subgroup of permutations. However, many DFAs have transition monoids which do not contain ``interesting'' subgroups; for example, DFAs of star-free languages have transition monoids with no non-trivial subgroups. In these cases, what can we say about uniform boolean minimality, or even accessibility of direct products?
\item
Look for necessary and/or sufficient conditions characterizing state complexity witnesses for operations on regular languages other than boolean operations (e.g., concatenation, star, reverse). As these problems could be extremely difficult, it may be useful to start with the special case of group languages or some other subclass of the regular languages.
\ei

\subsection*{Acknowledgements}
I thank Jason Bell and Janusz Brzozowski for careful proofreading and helpful comments. The computer algebra system GAP~\cite{GAP} was invaluable for this research; I cannot overstate its importance in obtaining these results. In particular, I thank the authors of the Automata GAP package~\cite{GAPAut} and all contributors to GAP's library of primitive groups.

\e{Funding:} This work was supported by the Natural Sciences and Engineering Research Council of Canada under grant No. OGP0000871.

\bibliographystyle{splncs03}
\bibliography{groups}

\end{document}